\documentclass[twocolumn,pra,aps,superscriptaddress,showpacs]{revtex4}
\usepackage{xspace,amsmath,amsfonts,amsthm,amssymb,amsbsy}
\usepackage{graphicx}
\usepackage{epstopdf}
\usepackage[normalem]{ulem}	% Part of the standard distribution
\usepackage{bbm}
\usepackage[breaklinks=true]{hyperref}

\newcommand{\smallfrac}[2]{\mbox{$\frac{#1}{#2}$}}
\newcommand{\half}{\smallfrac{1}{2}}
\newcommand{\bra}[1]{\langle{#1}|}
\newcommand{\ket}[1]{|{#1}\rangle}

\newcommand{\op}[2]{\ket{#1}\bra{#2}}

\newcommand{\expt}[1]{\langle{#1}\rangle}
\newcommand{\dg}{^\dagger}

\newcommand{\emn}[1]{ \mathbbm{E}_{#1} }

\newcommand{\beq}{\begin{equation}}
\newcommand{\eeq}{\end{equation}}
\newcommand{\bqa}{\begin{eqnarray}}
\newcommand{\eqa}{\end{eqnarray}}
\newcommand{\nn}{\nonumber}

\newcommand{\erf}[1]{Eq.~(\ref{#1})}
\newcommand{\frf}[1]{Fig.~\ref{#1}}
\newcommand{\srf}[1]{Sec.~\ref{#1}}
\newcommand{\sch}{Schr\"odinger}

\newcommand{\Tr}{\mbox{Tr}}

\newcommand{\str}{Stratonovich }



\begin{document}

\newtheorem{theo}{Theorem}
\newtheorem{lemma}{Lemma}
%@twocolumnfalse\endcsname

\title{$N$-Photon wave packets interacting with an arbitrary quantum system}
\author{{Ben Q. Baragiola}}
\affiliation{Center for Quantum Information and Control, University of New Mexico, Albuquerque, NM 87131-0001, USA}
\author{{Robert L. Cook}}
\affiliation{Center for Quantum Information and Control, University of New Mexico, Albuquerque, NM 87131-0001, USA}
\author{{Agata M. Bra\'nczyk}}
\affiliation{Department of Physics and Centre for Quantum Information and Quantum Control, University of Toronto, Toronto ON M5S 1A7, Canada}
\author{{Joshua Combes}}
\affiliation{Center for Quantum Information and Control, University of New Mexico, Albuquerque, NM 87131-0001, USA}

\begin{abstract}
	We present a theoretical framework that describes a wave packet of light prepared in a state of definite photon number interacting with an arbitrary quantum system (e.g. a quantum harmonic oscillator or a multi-level atom). Within this framework we derive master equations for the system as well as for output field quantities such as quadratures and photon flux. These results are then generalized to wave packets with arbitrary spectral distribution functions.  Finally, we obtain master equations and output field quantities for systems interacting with wave packets in multiple spatial and/or polarization modes.
\end{abstract}

\pacs{03.67.-a,42.50.Ct, 42.50.Lc, 03.65.Yz}
% Quantum information, 03.67.-a
% Light interaction with matter, 42.50.Ct
% Nonlinear optics, 42.65.-k
% Quantum noise, 42.50.Lc
% Fluctuation phenomena quantum optics, 42.50.Lc
% Quantum optics, 42.50.-p
% Antibunched photon states, 42.50.Dv
%    03.65.Yz - Decoherence; open systems; quantum statistical methods

\maketitle

%%% SECTION:  INTRODUCTION AND MOTIVATION %%%

	Nonclassical states of light are important resources for quantum metrology \cite{GioLor11, Leroux11}, secure communication \cite{BevGrangier02}, quantum networks \cite{Kimble08, MoeMauOlm07, Agh11}, and quantum information processing \cite{Nielsen05,KLM01}. Of particular interest for these applications are traveling wave packets prepared with a definite number of photons in a continuous temporal mode, known as \emph{continuous-mode Fock states} \cite{BlowLouden90,Lou_book00,GarChi_book08,Roh07}. As the generation of such states becomes technologically feasible \cite{BullColl10, Varcoe04, Belthangady09, Yamamoto06,  Zeilinger06, Walmsley08, Silberberg10, Kuhn11, LeePatPar12,McKBocBoo04,SpecBocMuc09,KolBelDu08} a theoretical description of the light-matter interaction \cite{GarPar94} becomes essential, see \frf{fig1}.

	Previously, aspects of continuous-mode single-photon states interacting with a two-level atom have been examined. Others have investigated master equations \cite{GheEllPelZol99}; two-time correlation functions \cite{GheEllPelZol99,DomHorRit02}; properties of scattered light \cite{DroHavBuz00, DomHorRit02, SheFan05,Kos08,CheWubMor11,ZheGauBar10,SheFan07b,ZhoGonSun08,LonSchBus09,Roy11,Roy10,Ely12}; and optimal pulse shaping for excitation \cite{StoAlbLeu07,WanSheSca10,StoAlbLeu10,RepSheFan10,Ely12}.  The results in these studies were produced with a variety of methods which have not been applied to many systems other than two-level atoms or Fock states where $N \gg 1$, however see~\cite{YudRei08}.

	One way to approach such problems is through the input-output formalism of Gardiner and Collett \cite{ColGar84,GarCol85,GarZol00,YurDen84,Caves82}. A central result of input-output theory is the Heisenberg-Langevin equation of motion driven by \emph{quantum noise} that originates from the continuum of harmonic oscillator field modes \cite{GarZol00,qnoise_rmp10}. The application of input-output theory to open quantum systems has historically been restricted to Gaussian fields \cite{GarCol85,DumParZol92,GarZol00} ---vacuum, coherent, thermal, and squeezed--- with several notable exceptions \cite{Gar93, Car93,GheEllPelZol99,GJNphoton,GJNCgen}.

%%% FIGURE 1:  Schematic %%%

	\begin{figure}[ht!]
    		\begin{center}
    		\includegraphics[width=1\hsize]{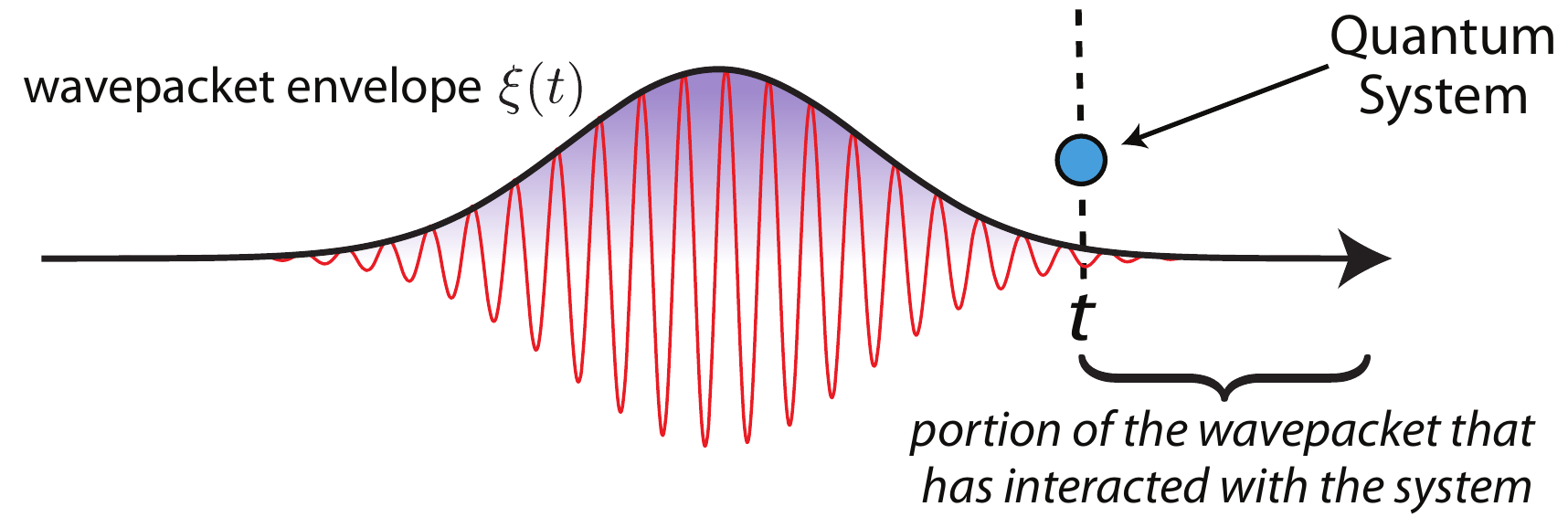}
       		 \caption{(Color online) Schematic depiction of a traveling wave packet interacting with an arbitrary quantum system.  The temporal wave packet is described by a slowly-varying envelope $\xi(t)$ which modulates fast oscillations at the carrier frequency.  We consider the case where the wave packet is prepared in a nonclassical state of definite photon number.} \label{fig1}
    		\end{center}
	\end{figure}

	In this article we present a unifying method, based on input-output theory, for describing the interaction between a quantum system and a continuous-mode Fock state. Consequently our formalism encapsulates and extends previous results. Specifically our method allows one to derive the master equations and output field quantities for an arbitrary quantum system interacting with any combination of continuous-mode $N$-photon Fock states. 

	This article is organized as follows: In \srf{sec_techintro} we introduce the white noise Langevin equations of motion, the mathematical description of quantum white noise, and the formal definition of continuous-mode Fock states.  In \srf{sec_fock} we present the first main results: the method for deriving master equations for systems interacting with continuous-mode Fock states and related output field equations.  This result is then extended in \srf{Sec::GeneralWavepackets} to continuous-mode ``$N$-photon states," where the spectral density function is not factorizable.  Then, in \srf{Sec::2LevelExcitation} we apply our formalism to the study of a two-level atom interacting with wave packets prepared in $N$-photon Fock states.  This application is intended to serve as an instructive example that reproduces and extends results in previous studies \cite{StoAlbLeu07,WanSheSca10,StoAlbLeu10}.  In \srf{sec_2modefock}, we present the second main result: master equations and output field quantities for a system interacting with Fock state wave packets in two modes (e.g. spatial or polarization). This sets the stage for the study of many canonical problems in quantum optics. As a two-mode example, we examine the scattering of Fock states from a two-level atom in \srf{Sec::2modeExample}. Finally, we conclude in \srf{sec_dis} with discussion and possible applications.

%%% SECTION:  Model and Methods %%%
\section{Model and Methods} \label{sec_techintro}

	A description of a system interacting with a traveling wave packet naturally calls for a formulation in the time domain.  The input-output theory developed in the quantum optics community provides such a description \cite{GarCol85,YurDen84,Caves82,DumParZol92,GarZol00,Gar93, Car93}.  Often input-ouput theory is formulated for a one-dimensional electromagnetic field, although this is not a necessary restriction \cite{DumParZol92}. (Such effective one-dimensional models are typically thought about in the context of optical cavities \cite{Aoki09} or photonic waveguides \cite{CheWubMor11,Spillane08, Vetsch10,Chang07}.) In this formalism the rotating wave approximation, the weak-coupling limit (the Born approximation), and the Markov approximation are made \cite{VanHove55,vanHStoMab05a}.  Strict enforcement of these approximations is known as the \emph{quantum white noise limit} \cite{Accardi}.  

	In Appendix (\ref{SEC::QWNL}) we review the quantum white noise limit; other introductory material can be found in Refs.~\cite{GarCol85,GarZol00,YanKim03a, vanHStoMab05a}. The main result is a quantum stochastic differential equation (QSDE) for the unitary time evolution operator that governs the system-field dynamics. From this equation one can derive QSDEs for system and field operators driven by white noise, also known as white-noise Langevin equations.  It is these equations of motion that lie at the heart of the derivation of Fock-state master equations.

	The Langevin equations derived in the white noise limit are in \str form \cite{GarZol00,Louisell1973book,GarChi_book08}.  \str QSDEs obey the standard rules of calculus, but expectations can be hard to calculate because the quantum noises do not commute with the operators to which they couple. \str QSDEs can be converted to an equivalent form known as the It\={o} QSDEs.  In It\={o} form the quantum noises commute with the operators to which they couple, which facilitates taking expectations. However, differentials must be calculated to second order \cite{GarZol00}.  To derive master equations we will be taking expectations over field states and consequently will work solely with It\={o} QSDEs.

	%% SUBSECTION: Equations of Motion %%
	\subsection{Derivation of the vacuum master equation from the It\={o} Langevin equations }	

	Consider an arbitrary system operator in the interaction picture, $X(t)$, with the initial condition $X(t_0)=X \otimes I_{\rm field}$.  The time evolution of $X$ is given by the It\={o} Langevin equation [see Appendix (\ref{App::ItoLangEQS})]
	\begin{align} \label{Eq::ItoLangevin}
		\nn dX =&( i[H, X] + \mathcal{L}\dg[L]X) dt  \\
		& +[L\dg,X]S dB_{t} + S\dg[X,L] dB\dg_{t} \\
		&+ (S\dg XS-X) d\Lambda_{t}, \nn
	\end{align}
where the action of the superoperator is 
	\begin{equation} \label{Eq::LinbladHeiO}
		\mathcal{L}\dg[L]X = L\dg XL- \half \left( L\dg L X + X L\dg L  \right).
	\end{equation}
	The operators $(S,L,H)$ act on the system Hilbert space. The quantum noise increments $dB_t$, $dB_t\phantom{}\dg$, and $d\Lambda_t$ are field operators, discussed in more detail shortly. 

	 The first two terms in \erf{Eq::ItoLangevin} describe smooth evolution from an external Hamiltonian on the system and from a Lindblad-type dissipator.  The second two terms describe the influence of quantum noise through coupling of a system operator $L$ linearly to the field operators, e.g. dipole-type coupling. The final term arises from coupling of a system operator $S$ to a quantity quadratic in the field operators, such as photon number. Such effective couplings appear in optomechanical systems \cite{Van11} and arise after adiabatic elimination of the excited states in multi-level atoms \cite{DeuJes09}, for example.

	Let us return to the discussion about the quantum noise increments $dB_t$, $dB_t\phantom{}\dg$, and $d\Lambda_t$. These field operators are defined in terms of the fundamental field operators $b(t)$ and $b\dg(t)$ whose time arguments are mode labels rather than indicators of time evolution.
They are often referred to as white noise operators because they satisfy the singular commutation relations $[b(s), b\dg(t)]=\delta(t-s)$.  This is akin to classical white noise which is $\delta$-correlated in time.  Due to the singular nature of $b(t)$ and $b\dg(t)$, it is preferable to work with the quantum noise increments: 
	\begin{align}
		&dB_t = \int_t^{t+dt} ds \, b(s) \quad \text{and} \quad dB^\dagger_t = \int_t^{t+dt} ds \, b^\dagger(s),\\
		&\hspace{2cm}	d\Lambda_t = \int_t^{t+dt} ds \, b\dg(s) b(s), 
	\end{align}
which drive the Heisenberg dynamics in \erf{Eq::ItoLangevin}.

	Under vacuum expectation, the calculus rules for manipulating QSDEs are summarized by the relations
	\begin{align} \label{Eq::ItoTable}
		\begin{array}{l}
		dB_t dB\dg_t = dt,\,\,\, dB_t d\Lambda_t = dB_t, \\
		d\Lambda_t
		d\Lambda_t = d\Lambda_t,\,\,\, d\Lambda_t dB\dg_t= dB\dg_t.
		\end{array}
	\end{align}
These composition rules are often referred to as the \emph{vacuum It\={o} table.}

	As a prelude to the derivation of the Fock-state master equations, we derive the vacuum master equation.  First, we take vacuum expectations of \erf{Eq::ItoLangevin} using the following notation (to be explained in \srf{sec_fock}):  $\mathbb{E}_{0,0}[dX]=\Tr [ (\rho_{\rm sys} \otimes \op{0}{0})\dg dX]$. Consequently, we need the action of the quantum noise increments on vacuum,
	\begin{align}
		dB_t\ket{0}&=0,\\
		d\Lambda_t\ket{0}&=0.
	\end{align}
All of the quantum noise terms in \erf{Eq::ItoLangevin} vanish under vacuum. Then, using the cyclic property of the trace we obtain the vacuum master equation:
	\begin{align}\label{Eq::vacME}
		\frac{d}{dt} \varrho_{0,0}(t) = - i[H, \varrho_{0,0} ] + \mathcal{L}[L] \varrho_{0,0},
	\end{align}
where the Lindblad superoperator is defined as
	\begin{align}  \label{Eq::LindbladSuperO}
		\mathcal{L}[L] \varrho = L \varrho L\dg - \half \left( L\dg L \varrho + \varrho L\dg L \right)\!,
	\end{align}
and the subscripts on $\varrho_{0,0}$ denote that \erf{Eq::vacME} is a vacuum master equation.

	% SUBSECTION: Continuous-Mode Fock States %
	\subsection{Continuous-mode Fock states}\label{SEC:mulitphoton}

	A continuous-mode single-photon state \cite{Lou_book00,GarChi_book08,BlowLouden90} can be interpreted as a single photon coherently superposed over many spectral modes \cite{Mil07,Mil08} with weighting given by the spectral density function (SDF) $\tilde{\xi}(\omega)$,
	\begin{align}  \label{Eq::SinglePhotonFreqDomain}
		\ket{1_{\xi}} & = \int d\omega \, \tilde\xi(\omega) b\dg(\omega)\ket{0}.
	\end{align}
We focus on quasi-monochromatic wave packets, where the spectral spread is much smaller than the carrier frequency, $\Delta \omega \ll \omega_c$ \footnote{The time domain condition for an envelope to be slowly varying is $|\partial^2 \xi(t) / \partial t^2|\ll w_c |\partial \xi(t) / \partial t | \ll w_c ^2| \xi(t) |$ \cite{GarChi_book08}.}.  This holds for optical carriers, whose bandwidths are small relative to the carrier frequency.  Then, we can define a \emph{slowly-varying envelope} $\tilde\xi(\omega)$ rotating at the carrier frequency,
	\begin{align}
		\tilde\xi(\omega) \rightarrow \tilde \xi(\omega)e^{-i \omega_c t},
	\end{align}
where $\omega_c$ is near any relevant system frequencies.  The Fourier transform of the slowly-varying envelope, $\mathcal{F}[\tilde{\xi}(\omega)] = \xi(t)$, characterizes a square-normalized temporal wave packet, $\int dt \, |\xi(t)|^2 = 1$.  In the time domain, and within the quasi-monochromatic approximation, the single-photon state in Eq. (\ref{Eq::SinglePhotonFreqDomain}) becomes \cite{Lou_book00},
	\begin{align}\label{eqBdef}
		\ket{1_{\xi}} & = \int ds \, \xi(s) b\dg(s)\ket{0} \nonumber \\
		& \equiv B\dg (\xi) \ket{0},
	\end{align} 
where we have absorbed the possible detuning from the system frequency into $\xi(t)$. The operator $B\dg (\xi)$ creates a single photon in the wave packet $\xi(t)$.  Equation (\ref{eqBdef}) can be interpreted as a superposition of instantaneous photon creation times weighted by the temporal wave packet. Since the white noise operators are defined in the interaction picture, it is clear that $\xi(t)$ is a slowly-varying temporal envelope rotating at the carrier frequency. By focusing on quasi-monochromatic wave packets we ensure the approximations made in the quantum white noise limit are not violated.

	A straightforward extension leads to the definition of normalized, continuous-mode Fock states (referred to hereafter as \emph{Fock states}) in the wave packet $\xi(t)$ with $N$ photons \cite{BlowLouden90},
	\begin{subequations} \label{Eq::ContModeFockState}
	\begin{align}
		\ket{N_{\xi}}  &=\frac{1}{\sqrt{N!}}\left [\int ds \, \xi(s) b\dg(s)\right]^{N} \ket{0}\\
                   &=\frac{1}{\sqrt{N!}}\left [ B\dg(\xi)\right ]^{N}\ket{0}.
	\end{align}
	\end{subequations}
The Fock states in \erf{Eq::ContModeFockState} are a subset of more general $N$-photon states for which the SDF is not factorizable ~\cite{Roh07}.  In \srf{Sec::GeneralWavepackets}, we define these states and use them to derive master equations.

%%% SECTION: Fock State Master Equations %%%
\section{Fock State Master Equations} \label{sec_fock}

	In this section we derive master equations for a quantum system interacting with a field prepared in a Fock state.  The derivation is performed in the interaction picture where the time-dependent operators evolve according to \erf{Eq::ItoLangevin}. To facilitate the derivation we first introduce notation convenient for representing expectations with respect to a particular field state. It should be noted that our method is a generalization to $N$-photon states of a method introduced in Refs.~\cite{GJNphoton,GJNCgen}  for a single photon.

	Assuming no correlations before the interaction, the total system is described by the product state 
	\begin{align}\label{Eq::initState}
		\rho(t_0) = \rho_{\rm sys} \otimes \op{N_\xi}{N_\xi}, 
	\end{align}
with the system in the state $\rho_{\rm sys}$ and the field in the Fock state $\ket{N_\xi}$. Using the Hilbert-Schmidt inner product for operators $A$ and $B$,
	\begin{align} \label{Eq::HSInnerP}
		\expt{A|B} \equiv \Tr [A\dg B],
	\end{align}
one can take expectations with respect to system and/or field states. For the following derivation it is necessary to define the asymmetric expectation value,
	\begin{align}\label{Eq::GenExpectation1}
		\mathbbm{E}_{m,n}[O] \equiv \Tr_{\rm sys + field} \left[ \left( \rho_{\rm sys} \otimes \op{m_\xi}{n_\xi} \right) \dg O \right]
	\end{align}
where $O$ is a {\em joint} operator on the system and field and is not necessarily separable.  We use a convention where capital letters, $\ket{N_\xi}$ denote the number of photons in the input field. Lowercase letters, that is, $\ket{n_\xi}$ where $n = \{0,...,N\}$, label ``reference'' Fock states to which the system couples.  Using the Hilbert-Schmidt inner product, we define a set of generalized density operators $\varrho_{m,n}$, first introduced in Ref.~\cite{GheEllPelZol99}, by tracing over only the field in \erf{Eq::GenExpectation1}:
	\begin{align}  \label{Eq::GeneralDensityOps}
		\mathbbm{E}_{m,n}[ O ] &  \equiv  \Tr_{\rm sys }[ \varrho\dg_{m,n} O].
	\end{align}
Such generalized density operators were also used in Refs.~\cite{GJNphoton,GJNCgen}  for a single photon. We delay the interpretation of these generalized density operators until \srf{sec_fockgen}.

	As the trace in \erf{Eq::GenExpectation1} is over both system and field, it gives a $c$-number expectation value. 
Using the partial trace we also define an asymmetric partial expectation over the field alone which results in an operator.  We define this operation with the notation \footnote{Note that the symbol $\varpi$ was used in Refs.~\cite{GJNphoton,GJNCgen}. Our definition is different.},
	\begin{align} \label{Eq::GenExpectation2}
		\varpi_{m,n}( O ) &\equiv \Tr_{\rm field} \left[ \left( I_{\rm sys}\otimes  \op{m_\xi}{n_\xi} \right) \dg O  \right]\!.
	\end{align}
We base our derivation on the It\={o} Langevin equations of motion for system operators.  In this picture, the state remains separable and the expectations will always have the form of \erf{Eq::GenExpectation1} and \erf{Eq::GenExpectation2}.  

	At this point we must mention an important technical issue.  The composition rules for the quantum noise increments, expressed in \erf{Eq::ItoTable}, are generally modified for non-vacuum fields \cite{GarZol00,WisMilBook}.  {However, it is shown in Appendix \ref{itotable}  that the It\={o} table for Fock states is identical to that for vacuum.  This allows the techniques from input-output theory to be extended to Fock states.}

	%% SUBSECTION: General Fock State Master Equation %%
	\subsection{Fock-state master equations for the system }\label{sec_fockgen}

	Recall the first step towards deriving the vacuum master equation, \erf{Eq::vacME}, was taking the expectation of \erf{Eq::ItoLangevin} with respect to vacuum, i.e. $\mathbbm{E}_{0,0}[dX]$.  Analogously, to derive the Fock-state master equations we must take the asymmetric expectations, i.e. \erf{Eq::GenExpectation1} or \erf{Eq::GenExpectation2}. The only explicit field operators in \erf{Eq::ItoLangevin} are the quantum noise increments $dB_{t}$ and $d\Lambda_{t}$. Consequently the action of the quantum noise increments on Fock states is needed:
	\begin{subequations} \label{Eq::IncrementActions}
	\begin{align}
		dB_{t}\ket{n_{\xi}}& = dt\sqrt{n} \xi(t) \ket{n-1_\xi}, \label{second} \\
		d\Lambda_{t}\ket{n_{\xi}}& = dB\dg_t \sqrt{n} \xi(t) \ket{n-1_\xi}. \label{third}
	\end{align}
	\end{subequations}	  
In Appendix \ref{qnoise_inc} we show how to derive these relations. Equations (\ref{Eq::IncrementActions}) show how ``reference" Fock states of different photon number couple through the quantum noise increments. 

	We are now equipped to derive the Fock-state master equations. From \erf{Eq::GenExpectation2}, we take the partial trace over Fock states for an arbitrary system operator $X \otimes I_{\rm field}$, whose equation of motion is given by \erf{Eq::ItoLangevin}. Doing so yields the Heisenberg master equations:
	\begin{align} \label{Eq::SingleModeMEHei}
		 \frac{ d }{dt} \varpi_{ m,n}( X(t) ) = & \varpi_{m,n}(i[ H, X ] ) + \varpi_{m,n}(\mathcal{L}\dg[ L] X) \\
          	&+ \nn \sqrt{m} \xi^*(t) \varpi_{m-1,n}( S\dg [X, L])\\
          	&+ \nn \sqrt{n} \xi(t) \varpi_{m,n-1}([L\dg,X]S) \\
          	&+ \nn \sqrt{ m n } |\xi(t)|^{2} \varpi_{m-1,n-1}(S\dg XS-X). 
	\end{align}

	To extract the \sch-picture master equations, we make use of \erf{Eq::GeneralDensityOps}: $\mathbbm{E}_{m,n}[ X(t) ] =   \Tr_{\rm sys }[ \varrho\dg_{m,n}(t) X]$. Then, using the cyclic property of the trace, we can write down the master equations for the system state:
	\begin{align} \label{Eq::SingleModeMESch}
 		\frac{d}{dt} \varrho_{m,n}(t) & = - i[H, \varrho_{m,n} ] + \mathcal{L}[L] \varrho_{m,n}   \\
 		& \!+\! \sqrt{m} \xi(t) [S \varrho_{m-1,n}, L^\dagger] \!+\!  \sqrt{n} \xi^*\!(t) [L,\varrho_{m,n-1} S^\dagger  ] \nonumber \\
		& \!+\! \sqrt{mn} |\xi(t)|^2\!\left(S \varrho_{m-1,n-1} S^\dagger - \varrho_{m-1,n-1}  \right) \nonumber\!.
	\end{align}
This set of coupled differential equations is the main result of this section. The initial conditions for these equations are: the diagonal equations $\varrho_{n,n}$ should be initialized with the initial system state $\rho_{\rm sys}$, while the off-diagonal equations should be initialized to zero. In order to calculate expectation values of system operators for an $N$-photon Fock state one needs only the top-level density operator $\varrho_{N,N}$.  However, extracting $\varrho_{N,N}$ requires propagating all equations between 0 and $N$ to which it is coupled. We note some special cases of \erf{Eq::SingleModeMESch} have been derived previously in Refs.~\cite{GheEllPelZol99,GJNphoton,GJNCgen} however little intuition or physical interpretation was given to these equations.

	The master equations in \erf{Eq::SingleModeMESch} require further explanation. The diagonal terms, $\varrho_{n,n}$, are valid state matrices describing the evolution of the system interacting with an $n$-photon Fock state for $n \in \{ 0, \dots, N \}$.  For example, when $N=0$ we recover the vacuum master equation: $d \varrho_{0,0} = - i[H, \varrho_{0,0} ] dt +  \mathcal{L}[L] \varrho_{0,0} dt,$ which is the only closed-form equation in \erf{Eq::SingleModeMESch}.  For $N \geq 1$, the diagonal equations couple ``downward'' towards the vacuum master equation via the off-diagonal equations $\varrho_{m,n}$ where $m \neq n$.  These off-diagonal operators are non-Hermitian of trace-class zero \cite{GheEllPelZol99}; consequently they are not valid state matrices but do satisfy $\varrho_{m,n}=\varrho_{n,m}\dg$.

	The fact that the equations couple downward means that we need only consider a \emph{finite} set of equations, which can be integrated numerically and in some cases, analytically.  For a field in an $N$-photon Fock state there are $(N+1)^2$ equations.  From the symmetry $\varrho_{n,m}=\varrho_{m,n}\dg$, the number of independent coupled equations reduces to $\half(N+1)(N+2)$. 

	Finally, we comment on the physical interpretation of these equations. Absorption of a photon by the system significantly changes a field prepared in a Fock state, so its dynamics are non-Markovian \cite{GheEllPelZol99,GJNphoton}.  This necessitates propagating a set of coupled master equations.  (In contrast, for coherent states photons can be removed while leaving the field state unchanged and a single master equation suffices.)  Before the wave packet has interacted with the system $\xi(t)$ is zero and only the top level equation $\varrho_{N,N}$ contributes to the evolution of the system.  In other words, the system evolves solely under the terms on the first line of \erf{Eq::SingleModeMESch}, which describe evolution from an external Hamiltonian and decay due to coupling to the vacuum. When the wave packet begins to interact with the system, $\xi(t)$ becomes nonzero and the other coupled equations contribute to the evolution of the system. Then, the information flow propagates \emph{upwards} from $\varrho_{0,0}$ to $\varrho_{N,N}$ because the equations couple downwards. 
	
So far we have discussed the dynamics of the system before and during the interaction. The last physically important observation is related to the correlation between the system and the outgoing field during and {\em after} the interaction. Consider the case where $\xi(t)$ is bimodal.  When the temporal spacing between the peaks is much greater than the characteristic decay time of the system and since $\xi(t)$ is zero at these intermediate times, the coherence between the first peak of the wave packet and the system is lost before the second peak begins to interact. Thus only the top-level equation must be propagated at these times, and the only nonzero terms describe external Hamiltonian drive and decay into the vacuum.  When the temporal spacing between the two peaks is on the order of the system decay time or shorter,  then the initial temporal coherence between the peaks can affect the system.

	%% SUBSUBSECTION: Output Field Equations %%
	\subsection{Output field quantities }\label{Sec::outputfield}

	In addition to system observables, we may also be interested in features of the output field \footnote{Our formalism also applies to operators of the form $\mathfrak{O}(t_0)=X_{\rm sys}\otimes Y_{\rm field}$. Asymmetric expectations are taken as usual.}.  Consider a field observable $Y(t)$ with initial condition $Y(t_0) = I_{\rm sys} \otimes Y$.   We insert the It\=o Langevin equation of motion for $Y$ into the asymmetric expectations. Using \erf{Eq::GenExpectation2} for the partial trace $\varpi_{m,n}( Y(t) )$, the result is operator-valued Heisenberg master equations. We focus here on expectation values, $\mathbbm{E}_{m,n}[Y(t)]$, which are found by tracing over the system as well, as in \erf{Eq::GenExpectation1}.   For two field quantities of interest -- photon flux and field quadratures -- we produce a set of coupled differential equations similar in form to \erf{Eq::SingleModeMESch}.  The initial conditions are $\varpi_{m,n}(Y(t_0)) = 0 \cdot I$ and similarly $\mathbbm{E}_{m,n}[Y(t_0)] = 0$.

		%% SUBSUBSECTION:   Output photon flux %%
		\subsubsection{Photon flux}
		
	The photon flux is given by $d\Lambda_t$, which counts the number of photons in the field in the infinitessimal time increment $t$ to $t+ dt$ \cite[Sec. 11.3.1]{GarZol00}.  The rules of It\={o} calculus are used in Appendix \ref{SubSection::GeneralPropagator} to give the equation of motion for the output photon flux $\Lambda_t^{\rm out}$,
	\begin{align}
		d\Lambda_t^{\rm out} =   L\dg L dt  + L\dg S dB_t + S\dg L dB\dg_t + d\Lambda_t.
	\end{align}
Taking expectations over Fock states using \erf{Eq::GenExpectation1} yields an equation for the mean photon flux, 
	\begin{align} \label{Eq::SingleModeFieldME_lambda}
		\nn \frac{d}{dt}& \emn{m,n} [\Lambda^{\rm out}_t(t)] = \emn{m,n} [ L\dg L]  + \sqrt{m} \xi^*(t) \emn{m-1,n} [  S\dg L ]  \\
 		& + \sqrt{n} \xi(t) \emn{m,n-1} [ L\dg S ] + \sqrt{mn} |\xi(t)|^2  .
	\end{align}
The solution to this equation $\mathbbm{E}[\Lambda_t^{\rm out}(t)]$ gives the integrated mean photon number up to time $t$.

 		%% SUBSUBSECTION:  Output Quadratures %%
		\subsubsection{Field quadratures}	

	A Hermitian field quadrature $Z_t$ measurable via homodyne detection is described by  
	\begin{align}
		Z_t= e^{i \phi} B_t +e^{-i\phi} B\dg_t.
	\end{align}
Following the same prescription, the equation of motion for the quadrature after the interaction is 
	\begin{align}
		dZ_t^{\rm out}  &= e^{i\phi}dB_t^{\rm out} + e^{-i\phi}dB_t^{\dagger \rm out} \nonumber \\
		&= e^{i\phi}(L dt + S dB_t) + e^{-i\phi}(L\dg dt + S\dg dB_t\dg) .
	\end{align}
Taking expectations over Fock states using \erf{Eq::GenExpectation1} gives the mean homodyne current,
	\begin{align} \label{Eq::SingleModeFieldME_B}
		\frac{d}{dt}& \emn{m,n} [Z_t^{\rm out}(t)] = \emn{m,n}[e^{i \phi} L + e^{-i \phi} L\dg] \\
		&+ e^{i \phi} \sqrt{n} \xi(t) \emn{m,n-1}[ S ] + e^{-i \phi} \sqrt{m} \xi^*(t) \emn{m-1,n}[ S\dg ] \nonumber.
	\end{align}

	%% SUBSECTION: General Input States %%
	\subsection{General input field states in the same wave packet }\label{sec_fockinput}

	So far we have considered the case where the input field is a ``pure'' Fock state. These results can be generalized to field states described by an arbitrary combination (superposition and/or mixture) of Fock states in the same wave packet.  As the Fock states span the full Hilbert space, they form a basis for arbitrary states in the wave packet $\xi(t)$,
	\begin{align} \label{Eq::CominationStates}
		\rho_{\mathrm{field}} = \sum_{m,n=0}^\infty c_{m,n} \op{n_{\xi}}{m_{\xi}}.
	\end{align}
The coefficients are constrained by the requirements of valid quantum states:  $\rho _{\mathrm{field}} \geq 0$, $\mathrm{Tr}[\rho _{\mathrm{field}}]=1$ and $\rho _{\mathrm{field}\phantom{\dg}}=\rho _{\mathrm{field}}\dg$. 

	When the input field is described by \erf{Eq::CominationStates} the system state is
\begin{equation} \label{Eq::gen_me}
	\varrho_{\rm total} (t)=\sum_{m,n} c^*_{m,n} \varrho_{m,n}(t),
\end{equation}
where $\varrho _{m,n}(t)$ are the solutions to the master equations.  Generating the full, physical density operator for an arbitrary field requires combining the appropriate solutions from the hierarchy of coupled equations in \erf{Eq::SingleModeMESch} with associated weights $c_{m,n}$.  The Heisenberg master equation is found in the same manner,
\begin{equation} \label{Eq::GeneralHEM}
	\varpi_{\rm total} (t)=\sum_{m,n}c_{m,n} \varpi_{m,n}(t).
\end{equation}
Finally, the expectation value of a system operator $X$ is given by
	\begin{align}
		 \mathbbm{E}_{\rm total}[X(t)] &={\rm Tr}_{\rm sys+field}\left[ \varrho\dg_{\rm total}(t) X \right] \\
	&=\sum_{m,n} c_{m,n} \mathbbm{E}_{m,n}[X(t)] \label{Eq::GenExpectations}.
\end{align}
This technique also applies to the output field quantities in \srf{Sec::outputfield}.  Note that the definition of the Hilbert-Schmidt inner product, \erf{Eq::HSInnerP}, gives rise to the conjugate coefficients in Eq. (\ref{Eq::gen_me}) but not in Eqs. (\ref{Eq::GeneralHEM}, \ref{Eq::GenExpectations}).

%% SECTION: General Multi-photon Wavepacket Master Equations %%%
\section{General N-Photon Master Equations } \label{Sec::GeneralWavepackets}

	In many experimental settings multiple photons are not created in Fock states. Fock states are a subset of more general $N$-photon states, which have a definite number of photons but an arbitrary SDF $\tilde\psi(.)$.  Indeed, a quantum tomography protocol for characterizing the SDF was recently proposed \cite{RohdeSep06} and implemented \cite{WasKolRob07}.  This motivates the derivation of master equations for such fields.  
	
	In a single spatial and polarization mode, a general $N$-photon state is 
	\begin{align}\label{Eq::Genphotyfreq}
	\begin{split}
		\ket{\psi_N} ={}\int d\omega_1 & \dots d\omega_N \, \tilde\psi(\omega_1,\dots,\omega_N) \\
		&\times b\dg(\omega_1)\dots b\dg(\omega_N) \ket{0}\,.%\\
		\end{split}
	\end{align}  
Again we assume quasi-monochromatic wave packets such that $\psi(\cdot)$ is a slowly-varying envelope with respect to the carrier frequency.  Then, in the time domain a general $N$-photon state can be written as
	\begin{align}\label{Eq::Genphoty}
	\begin{split}
		\ket{\psi_N} ={}\int dt_1 & \dots dt_N \, \psi(t_1,\dots,t_N) \\
		&\times b\dg(t_1)\dots b\dg(t_N) \ket{0}\,.%\\
		\end{split}
	\end{align} 
These states are not amenable to our analysis directly. Thankfully, a formalism for dealing with such $N$-photon states has been developed~\cite{Roh07,Ou06}. 

	To describe $N$-photon states we make use of the occupation number representation developed by Rohde et al.~\cite{Roh07}, which we review in Appendix \ref{Appendix::multiphotonStates}.  Using \erf{AEq::NPhotonState}, \erf{Eq::Genphoty} can be written in a basis of orthogonal Fock states,
	\begin{align} \label{Eq::GeneralNPhotonState}
		\ket{ \psi_N } = \sum_{i_1\leq ... \leq i_N} \lambda_{i_1,\dots, i_N} \ket{ {n_1}_{\xi_1} } \ket{ {n_2}_{ \xi_2} } ...
	\end{align}
where $\ket{ {n_k}_{\xi_k} }$ is a normalized Fock state described by \erf{Eq::ContModeFockState} with $n_k$ photons in basis function $\xi_k(t)$.  Counting the number of subscripts on $\lambda$ in \erf{Eq::GeneralNPhotonState} gives the total number of photons $N$, and the value of any subscript $i_k$ reveals the basis function that photon is in.

	In order to derive the master equation, we must first write down the action of the quantum noise increments on \erf{Eq::GeneralNPhotonState}:
	\begin{align}\label{eq:bla}
		&dB_t \ket{\psi_N} = dt \sum_k \sqrt{n_k} \xi_k(t) \ket{\psi_{N-1}^k} \\
		&d\Lambda_t \ket{\psi_N} = dB^\dagger_t \sum_k \sqrt{n_k} \xi_k(t) \ket{\psi_{N-1}^k}
	\end{align}
where $\ket{\psi_{N-1}^k}$ is defined as
	\begin{align} \label{Eq::GenNPhotonMinusOne}
		\ket{\psi_{N-1}^k} \equiv \sum_{i_1\leq \dots \leq i_N} \lambda_{i_1, \dots , i_N} \ket{ {n_1}_{\xi_1} } ... \ket{ {n_k-1}_{\xi_k} } ...\,,
	\end{align}
and is interpreted to mean that a single photon in one of the basis Fock states has been annihilated.

	To derive the master equation for a system interacting with the field $\ket{\psi_N}$, an asymmetric expectation value needs to be defined for such states: $\mathbb{E}_{\psi_m,\psi_n}[O]= {\rm Tr}[(\rho_{\rm sys}\otimes \op{\psi_m}{\psi_n})\dg O]$. As before this defines the generalized density operators $\varrho_{\psi_m, \psi_n}$. Using these definitions the master equations for the generalized density operators are	
	\begin{align} \label{MultiPhotonMESch}
		\frac{d}{dt}  \varrho_{\psi_m, \psi_n }(t) =& \mathcal{L} [L] \varrho_{\psi_m, \psi_n }  - i [H,\varrho_{\psi_m, \psi_n } ] \\
		\nn& + \sum_k \sqrt{m_k} \xi_k (t) [S \varrho_{\psi_{m-1}^k, \psi_{n} }, \, L\dg]  \\
		\nn& + \sum_k \sqrt{n_k} \xi^*_k(t) [L, \, \varrho_{\psi_{m}, \psi_{n-1}^k } S\dg]  \\
		\nn& + \sum_{k,k'} \sqrt{m_k n_{k'} } \xi^*_k (t) \xi_{k'} (t) \\
		\nn&\;\;\;\,\,\, \times \big( S \varrho_{\psi^k_{m-1},\psi^{k'}_{n-1}} S\dg - \varrho_{ \psi^k_{m-1},\psi^{k'}_{n-1}} \big) .
	\end{align}
Each master equations couples to a set of equations enumerated by the indices $\{m,n, k\}$.  The total number of equations required to describe such a state depends on the overlap of the initial wave packet with the particular choice of basis.  Equations for the output field can also be derived for $N$-photon states, but we omit them for brevity. Equations similar to \erf{MultiPhotonMESch} were derived in Ref.~\cite{GheEllPelZol99} for two photons but did not include $d\Lambda_t$ or $S$.

	Finally, we can consider input fields in combinations (superpositions and/or mixtures) of different $N$-photon states. In particular we allow the total state to be a combination of different states with the same photon number {\em and} a combination of states with different photon numbers. To describe such a state first we need to consider a general combination of $N$-photon states. That is, 
	\begin{align}
		{\Psi}_{N} = \sum_{K,L\in\{ \psi, \phi .. \sigma\}} c_{L,K} \op{K_{N}}{L_{N}},
	\end{align}
where the summation is over different states with the same photon number $N$. Then we can sum over photon numbers to obtain the most general input field:
	\begin{align}\label{Eq::gen_spec_arb_n}
		\rho_{\mathrm{field}} = \sum_{p=0}^\infty \mathcal{C}_{p} \Psi _p
	\end{align}
The coefficients $c_{L,K}$ and $\mathcal{C}_p$ are constrained by the requirement that the input state be a valid quantum state. Using Eqs. (\ref{Eq::gen_spec_arb_n}) and (\ref{Eq::gen_me}), the equations for the system and output field can be found.

%% SECTION:  {Example: N=2 Fock State Master Equation}\label{sec_singlephotonME}
\section{ Example: Fock-state master equations for a two-level atom interacting with a Gaussian wave packet} \label{Sec::2LevelExcitation}	

	Efficient photon absorption is important for information transfer from a flying to a stationary qubit. In this section we analyze this problem with a study of the excitation probability and output field quantities for Fock states interacting with a two-level atom.  This problem has been studied before in much detail for a single photon in Refs.~\cite{StoAlbLeu07,WanSheSca10,StoAlbLeu10}. Our intention is to make a direct connection to established results and then to extend those results to higher photon numbers. Consequently, we do not focus on optimizing wave packet shapes as other studies have~\cite{StoAlbLeu07,WanSheSca10,StoAlbLeu10,RepSheFan10}.
	
	 The single-mode approximation in \srf{sec_techintro} is rooted in the presumption that the wave packet can be efficiently coupled to the two-level atom.  This has been considered in the case of a mode-matched wave packet covering the entirety of the $4\pi$ solid angle in free space \cite{StoAlbLeu07,WanSheSca10}.   A more widely applicable context is that of strongly confined 1D photonic waveguides \cite{RepSheFan10}.  In such systems the coupling rate into the guided modes $\Gamma_g$ can be much larger than into all other modes $\Gamma_\perp$, where the total spontaneous emission rate is $\Gamma = \Gamma_g + \Gamma_\perp$ ~\cite{CheWubMor11,Chang07}.  In the following analysis, we take the idealized limit that coupling to all other modes can be fully suppressed and we set $\Gamma_\perp = 0$.  To properly account for losses, a second mode can be introduced using the tools of \srf{sec_2modefock} and finally traced over.  
	
	In \srf{SEC::2photonME}, we examine the form of the master equation for the simple case of a two-photon Fock state. Next in \srf{SEC::1n2photons} we numerically examine a two-level atom interacting via a dipole Hamiltonian with a wave packet prepared with at most two photons. First we reproduce the single-photon excitation results from prior studies, then we broaden these results to include two photons and output field quantities. Finally in \srf{Sec::LargePhotonNumbers} we present a numerical study for large-photon-number Fock states. This allows us to explore the relationship between  excitation probability, bandwidth, interaction time, and photon number. For photon numbers $N \gg 1$, we identify a region of strong coupling. 

%% SUBSECTION: Two-photon Fock state %%
\subsection{Two-photon Fock state master equations}\label{SEC::2photonME}

%% FIGURE 2: 2 photon comparisons %%

	\begin{figure*}[t]
	\includegraphics[width=1\linewidth]{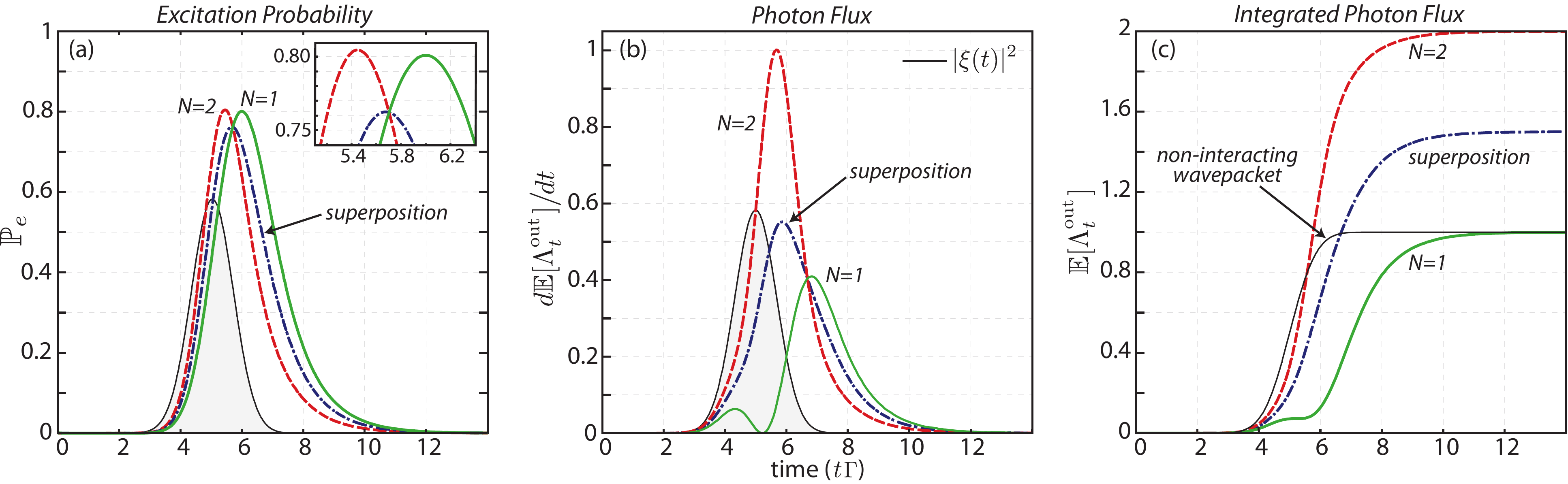} 
		\caption{(Color online) Comparison of a Gaussian wave packet of bandwidth $\Omega/\Gamma = 1.46$ in three initial field states: a single-photon Fock state (solid), a two-photon Fock state (dashed), and an equal superposition (dash-dot). The wave packet $|\xi(t)|^2$ is shown in black filled grey. (a) Excitation probability of a two-level atom. (b) Photon flux.  It is distinctly modified by interaction with the atom. (c) Integrated photon flux.  For comparison the integrated single-photon flux is plotted when there is no atom. \label{Fig::2PhotonCompare} }
	\end{figure*}

	It is instructive to examine the form of the master equation for the simple case of interaction with a two-photon Fock state where both photons are created in the same temporal wave packet $\xi(t)$, $\ket{\psi}_{ \mathrm{field}}= \ket{2_{\xi}}$. From \erf{Eq::SingleModeMESch}, the two-photon Fock state master equations are,
	\begin{widetext}
	\begin{subequations}\label{Eq::twophoton_eg}
	\begin{align}
 		\dot{\varrho}&_{2,2}(t) = \mathcal{L}[L] \varrho_{2,2} - i[H, \varrho_{2,2} ]  + \sqrt{2} \xi(t) [S \varrho_{1,2}, L^\dagger] +  \sqrt{2} \xi^*(t) [L,  \varrho_{2,1} S^\dagger ] + 2 |\xi(t)|^2\left( S \varrho_{1,1} S^\dagger - \varrho_{1,1}  \right) \label{rho22} \\
		\dot{\varrho}&_{2,1}(t) = \mathcal{L}[L] \varrho_{2,1} - i[H, \varrho_{2,1} ]
 		+ \sqrt{2} \xi(t) [S \varrho_{1,1}, L^\dagger] + \xi^*(t) [L, \varrho_{2,0} S^\dagger ]  \label{rho21} + \sqrt{2} |\xi(t)|^2\left( S \varrho_{1,0} S^\dagger - \varrho_{1,0}  \right)\\
		\dot{\varrho}&_{2,0}(t) = \mathcal{L}[L] \varrho_{2,0} - i[H, \varrho_{2,0} ]
 		+ \sqrt{2} \xi(t) [S \varrho_{1,0}, L^\dagger] \\ 	
		\dot{\varrho}&_{1,1}(t) = \mathcal{L}[L] \varrho_{1,1} - i[H, \varrho_{1,1} ]  + \xi(t) [S \varrho_{0,1}, L^\dagger] + \xi^*(t) [L, \varrho_{1,0} S^\dagger ]  \label{rho11} + |\xi(t)|^2\left(S \varrho_{0,0} S^\dagger - \varrho_{0,0}  \right)
\\
		\dot{\varrho}&_{1,0}(t) = \mathcal{L}[L] \varrho_{1,0} - i[H, \varrho_{1,0} ] + \xi(t) [S \varrho_{0,0}, L^\dagger] \label{rho10}\\
		\dot{\varrho}&_{0,0}(t) = \mathcal{L}[L] \varrho_{0,0} - i[H, \varrho_{0,0} ] \label{rho00}
	\end{align}
	\end{subequations}
	\end{widetext}
with the initial conditions:
	\begin{align}
		\varrho_{2,2}(0) &= \varrho_{1,1}(0) = \varrho_{0,0}(0) = \rho_{\rm sys} \\
		\varrho_{2,1}(0) &= \varrho_{2,0}(0) = \varrho_{1,0}(0) = 0.
	\end{align}
Similar equations to Eqs.~(\ref{Eq::twophoton_eg}) were originally derived in Ref.~\cite[Equations 71 (a)-(f)]{GheEllPelZol99} for a two-level atom but without the $S$ operator and the term proportional to $|\xi(t)|^2$. 
 For an arbitrary quantum system and single photon equations which include $S$ and the term proportional to $|\xi(t)|^2$ were later derived in Ref.~\cite{GJNphoton}. Then Ref.~\cite{GJNCgen} showed how to propagate these equations for any superposition or mixture of one photon and vacuum.

	Now suppose the input field is in a superposition of one and two photons, $\ket{\psi}_{\mathrm{field}} = \alpha \ket{1_\xi} + \beta \ket{2_\xi}$ with $ |\alpha|^2 + |\beta|^2 = 1$. From \erf{Eq::gen_me} we combine the solutions to the master equations, \erf{Eq::twophoton_eg}, to get the physical state,
	\begin{align}  \label{Eq::Varrho20}
		\varrho_{\rm total} (t) =& |\alpha|^2 \varrho_{1,1}(t) + |\beta|^2\varrho_{2,2}(t) \\
		\nonumber  & + \alpha^* \beta \varrho_{1,2}(t) + \alpha \beta^* \varrho_{2,1}(t). 
	\end{align}
Notice that the last two terms of \erf{Eq::Varrho20} originate in the coherences of the input field.  It is interesting that the ``off-diagonal,'' traceless, generalized density operators (e.g. $\varrho_{1,2}$) contribute to the calculation of physical quantities, albeit in Hermitian combinations. Had the field been a ``pure" Fock state or a statistical mixture of one and two photons, these terms would not appear. 

	Output field quantities are calculated in the same fashion as \erf{Eq::Varrho20}. For example, the mean photon flux is,
	\begin{align}
 		\mathbbm{E}_{\rm total} [\Lambda_{t}^{\rm out} (t)] &=  |\alpha|^2 \mathbbm{E}_{1,1} [\Lambda_t^{\rm out}] + |\beta|^2 \mathbbm{E}_{2,2} [\Lambda_t^{\rm out}] \nn \\ 
 		& + \alpha^* \beta \mathbbm{E}_{2,1} [\Lambda_t^{\rm out}] + \alpha \beta^* \mathbbm{E}_{1,2}[\Lambda_t^{\rm out}],
	 \end{align}
where \erf{Eq::GenExpectations} was used to calculate $ \mathbbm{E}_{\rm total} [.] $.

	%% SUBSECTION:  Excitation of a two level atom by a two-photon wavepacket
	\subsection{A two-level atom interacting with one- and two-photon Gaussian wave packets}\label{SEC::1n2photons}

	Now we specialize to a wave packet prepared with up to two photons interacting on a dipole transition with a two-level atom initially in the ground state $\ket{g}$.  In the absence of an external system Hamiltonian the master equation parameters are: $H=0$, $L= \sqrt{\Gamma} \op{g}{e}$, $S=I$, and the coupling rate is chosen for simplicity to be $\Gamma = 1$. We focus on a square-normalized Gaussian wave packet, as defined in Ref. \cite{WanSheSca10}, whose peak arrives at time $t_a$,
	\begin{align}\label{Eq::gau_xi}
		\xi_{\rm gau}(t)=\left( \frac{\Omega^2}{2 \pi} \right)^{1/4} \exp{\left[ -\frac{\Omega^2}{4} (t - t_a )^ 2\right] } ,
	\end{align}
with no detuning and frequency bandwidth $\Omega$.  For Gaussian wave packets the simple relationship between bandwidth and temporal width enables us to explore the tradeoff between interaction time and spectral support around resonance \footnote{ As defined in \erf{Eq::gau_xi}, the variance of $|\xi(t)|^2$ is $1/\Omega^2$ and of $\xi(t)$ is $\sigma_T^2 = 2/\Omega^2$.  The variance of $|\xi(\omega)|^2$ is $\Omega^2$ and of $\xi(\omega)$ is $\sigma_\omega^2 = \Omega^2/2$.  This parameterization of a Gaussian was chosen to aid comparison with previous studies.}. 

	To study the excitation probability we numerically integrate the master equations (\ref{rho22})--(\ref{rho00}).   Then, for a given input field state we calculate the excitation probability,
	\begin{align}
		\mathbbm{P}_e(t) & = \Tr \left[ \varrho_{\rm total}(t) \op{e}{e} \right], 
	\end{align} 
where $\varrho_{\rm total}$ is given by \erf{Eq::gen_me}.

	Figure \ref{Fig::2PhotonCompare}(a) presents the excitation probability for a two-level atom interacting with a Gaussian wave packet  \erf{Eq::gau_xi} prepared in a ``pure" Fock state of one and two photons as well as an equal superposition; $\alpha = \beta = 1/\sqrt{2}$ in \erf{Eq::Varrho20}. In the simulations we use a bandwidth known to be optimal for single-photon Gaussian wave packets: $\Omega/\Gamma = 1.46$ \cite{StoAlbLeu07}. This gives a maximum excitation probability of $\mathbbm{P}_e^{\rm max} \approx 0.801$ for $N=1$ as found in other works \cite{StoAlbLeu07,WanSheSca10,StoAlbLeu10}.  Putting a second photon in the wave packet slightly increases this to $\mathbbm{P}_e^{\rm max} \approx 0.805$; however, we see in \srf{Sec::LargePhotonNumbers} that this is not universal behavior for all bandwidths and photon numbers.

	In \frf{Fig::2PhotonCompare}(b) we plot the mean photon flux of the output field, $d\mathbbm{E}[\Lambda_t^{\rm out}]/dt$, after interaction with the atom.  For the single-photon wave packet, we see a drastic change in the output photon flux when the photon is being absorbed by the atom.  For two photons, however, much of the wave packet travels through the atom undisturbed, since a two-level atom can absorb at most one photon.  The related integrated mean photon flux, $\mathbbm{E}[\Lambda_t^{\rm out}]$, is plotted in Fig. \ref{Fig::2PhotonCompare}(c).  For these ``pure'' one- and two-photon Fock states there exist a definite number of excitations. Any excitation induced in the atom through absorption of a photon eventually decays back into the field. This is shown in Fig. \ref{Fig::2PhotonCompare}(c) where the integrated mean photon flux for long times approaches the number of initial excitations $\{1, 1.5, 2\}$.  During the absorption of the single-photon wave packet, the integrated intensity flattens out since the photon has been transferred to an atomic excitation and arrives only later after decay.  

	For a single-photon wave packet, the Schr\"{o}dinger equation can be solved analytically for the excitation probability \cite{StoAlbLeu10,CheWubMor11}:
	\begin{align} \label{Eq::AnalyticProb}
 		\mathbbm{P}_e(t) = e^{-\Gamma t} \left| \int_0^t dt'  \,  \xi(t') e^{-\frac{\Gamma}{2}t'}\right|^2.
	\end{align}
The simulations in \frf{Fig::2PhotonCompare} agree with the analytic expression in \erf{Eq::AnalyticProb}.  However, it is not clear that the method used to derive \erf{Eq::AnalyticProb} can be extended to higher photon numbers.

%  MATLAB CODE SECTION
%
%We have supplied sample MATLAB code using fourth-order Runge-Kutta methods in the supplementary material to enable the %reader to reproduce \frf{Fig::2PhotonCompare}. 

	%% SUBSECTION:  Higher photon numbers %%
	\subsection{Excitation for large photon numbers} \label{Sec::LargePhotonNumbers}

%% FIGURE 3:  Scaling Combo %%
	\begin{figure}
	\includegraphics[width=1\hsize]{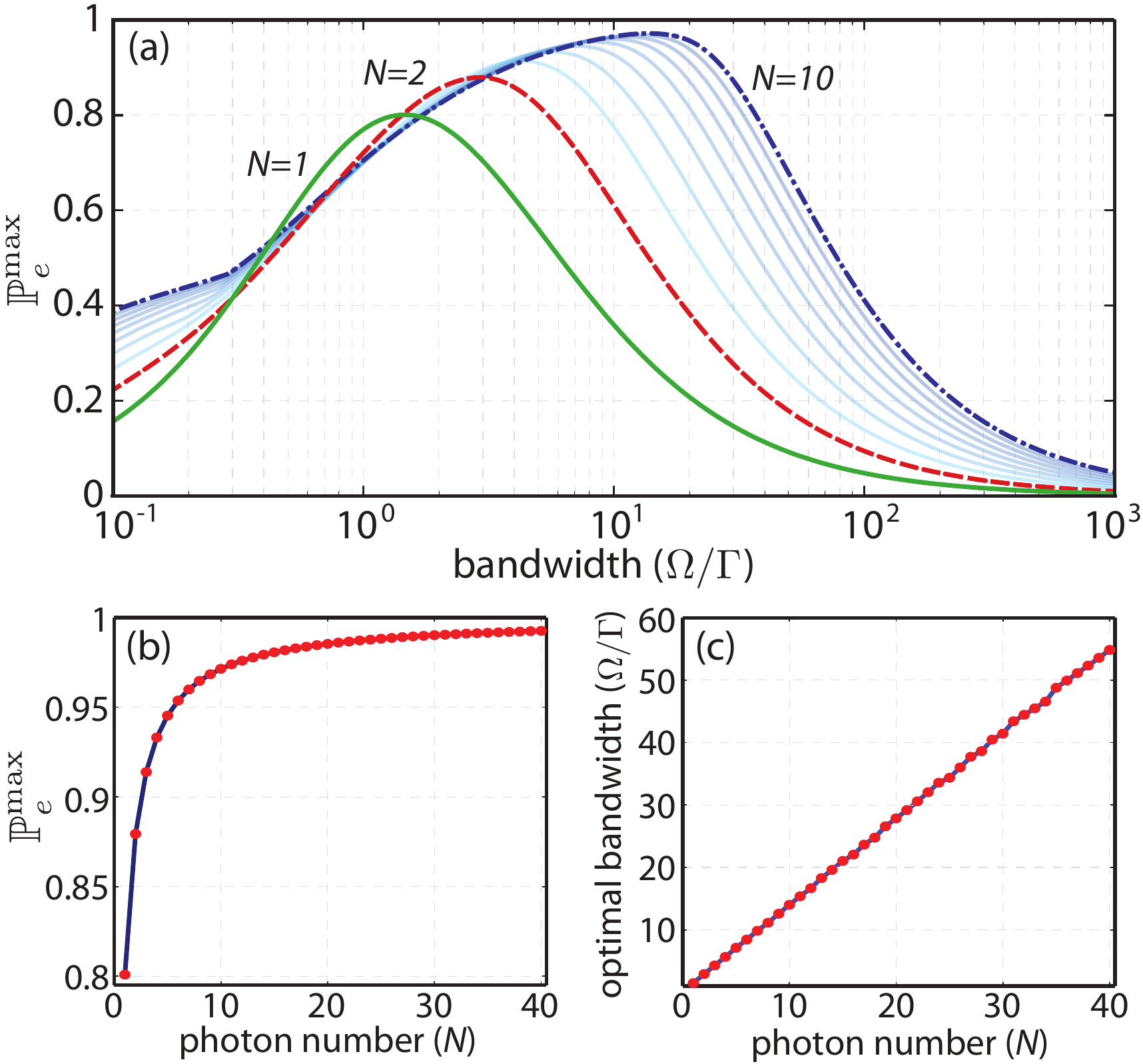}
		\caption{ (Color online) (a) Maximum excitation probability $\mathbbm{P}_e^{\rm max}$ of a two-level atom interacting with Gaussian wave packets of bandwidth $\Omega/\Gamma$ for photon numbers $N \in \{1,\dots 10\}$. Small (large) bandwidths correspond to long (short) temporal wave packets. (b) Scaling of $\mathbbm{P}_e^{\rm max}$ with photon number (red circles).  The fit shown is $\mathbbm{P}_e^{\rm max}(N) = 1-0.269\,N^{ -0.973}$ (blue line). (c ) Scaling of $\mathbbm{P}_e^{\rm max}$ with optimal bandwidth for each photon number $N$ (red circles).  The fit is $\Omega_{\rm opt} (N)/\Gamma = { 1.45} N^{0.987}$. Details of the fits can be found in the main text.
		} \label{Fig::NPhotonScaling} 
	\end{figure}
	In this section we expand the numerical study of excitation probability to Gaussian wave packets of the form of \erf{Eq::gau_xi} prepared Fock states with photon number $N\ge1$.

		%% SUBSUBSECTION:  Scaling %%
		\subsubsection{Scaling}
		
	For small bandwidths ($\Omega/\Gamma \ll 1$), see the left side of \frf{Fig::NPhotonScaling}(a), one would expect a high probability of excitation from the substantial spectral support near the transition frequency of the atom. However, the long temporal extent of the wave packet means the photon density over the relevant interaction time scale $\tau = 1/\Gamma$ is too small to significantly excite the atom \cite{WanSheSca10}. A complementary way of understanding this is that the dissipative terms in the master equations [terms on the first line of \erf{Eq::SingleModeMESch}] prevail over the coherent coupling (terms on the other lines). By extending the analysis in Ref.~\cite{GheEllPelZol99}, we find a recursive scaling of the excitation probability for very wide wave packets: $\mathbb{P}_{e}^{\rm max}\approx P_{N}$, where $P_{N} = N P_{1}(1-2P_{N-1})$  with  $P_{1}= 4\max |\xi(t)|^{2}$. 
	
  	In the other asymptotic regime where bandwidths are large ($\Omega/\Gamma \gg 1$), see the right side of \frf{Fig::NPhotonScaling}(a), the maximum excitation probability is small even for large photon numbers. This is due to the wave packet being so short that its bandwidth is spread over frequencies far from the atomic resonance. We numerically find the asymptotic scaling $\mathbb{P}_{e}^{\rm max}= 5 N\Gamma/\Omega$ for $\Omega/\Gamma \in [10^3,10^7]$ with $R^2=1$ for photon numbers $N \in \{1,\dots,10\}$. 
	
	At intermediate bandwidths, we note several interesting features.  First, the maximum excitation probabilities are not universally ordered by photon number and adding photons to the field can decrease $\mathbbm{P}_e^{\rm max}$.  In fact, there exists a bandwidth region in \frf{Fig::NPhotonScaling} where a single photon in the wave packet is optimal for excitation, $\Omega/\Gamma \approx [.5,1.4]$.
	
	Second, for each photon number there exists an optimal bandwidth for excitation.  In \frf{Fig::NPhotonScaling} (b) we have plotted the absolute maximum of $\mathbb{P}_{e}$ (maximized over $t$ and $\Omega/\Gamma$) as a function of the number of photons. We find excellent agreement ($R^2 = 1$) by fitting to the model $\mathbbm{P}_e^{\rm max}(N) = 1-aN^{-b}$ over the range $N \in \{10,\dots,40\}$  with coefficients (95\% confidence):  $a = 0.2694  (0.2678, 0.271), \,\,  b = 0.973  (0.9709, 0.975)$. Therefore the absolute maximum of $\mathbb{P}_{e}$ does monotonically increase with $N$, but with diminishing returns.
	
	In \frf{Fig::NPhotonScaling} (c) we investigate the optimal bandwidth for excitation for each photon number $N$. Fitting to the model $\Omega^{\rm max}(N)/\Gamma = aN^b$ gives $a=1.447(1.418, 1.476)$ and $b=0.9869(0.981, 0.9928)$ with 95\% confidence and $R^2=0.9998$.  Thus, to achieve this scaling for photon number $N$, the optimal bandwidth of the wave packet is $\Omega_{\rm opt} (N)/\Gamma \approx { 1.45} N^{0.987}$.  Thus, the optimal width seems to be proportional to the single-photon optimal bandwidth, $\Omega_{\rm opt} (N)/\Gamma \approx  1.46 N$.

		%% SUBSUBSECTION:  Dynamics %%
			\subsubsection{Dynamics}

\begin{figure*}
	\includegraphics[width=1\linewidth]{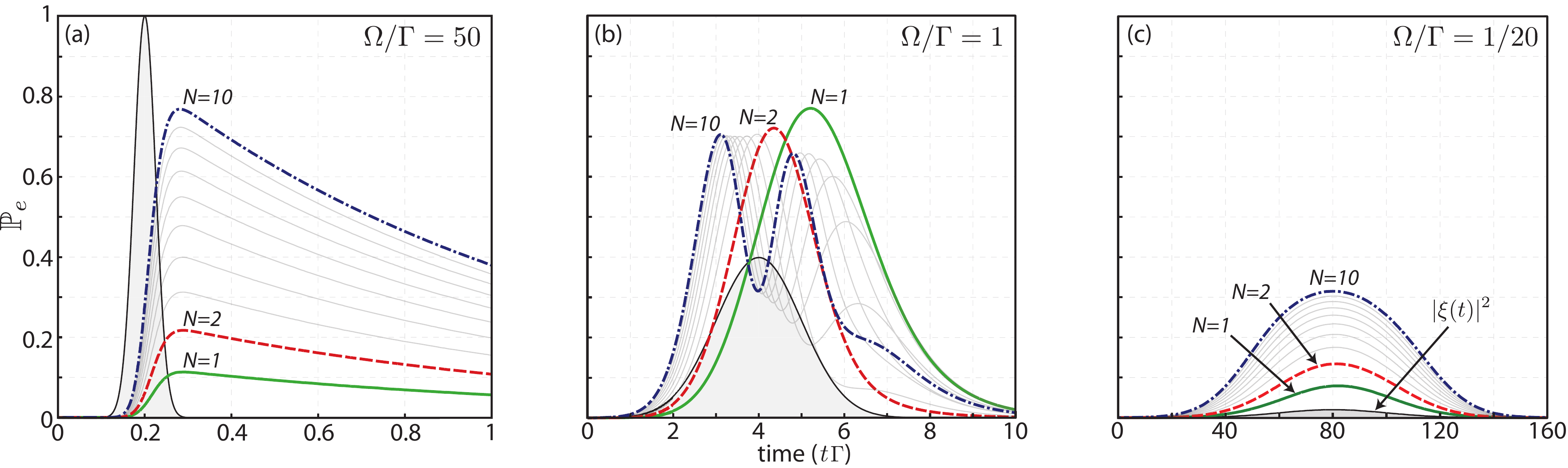}
	\vspace{-5mm} 
		\caption{ (Color online) Excitation probability $\mathbbm{P}_e$ of a two-level atom interacting with Gaussian wave packets of bandwidth $\Omega/\Gamma = \{ 50, 1, 1/20 \}$ prepared with $N \in \{1, \dots, 10\}$ photons. Highlighted are $N=1$ (solid), $N=2$ (dashed), and $N=10$ (dash-dot).  The wave packet $|\xi(t)|^2$ is plotted in black filled grey (normalized in (a) for clarity).  (a) Behavior of short temporal wave packets (large bandwidths) shows $\mathbbm{P}_e$ is ordered by photon number.  (b) For intermediate bandwidths, we see damped Rabi oscillations, discussed in \srf{SEC::StrongCoupling}.  Note that $\mathbbm{P}_e$ is not necessarily ordered. (c) Behavior of long temporal wave packets (small bandwidths) where $\mathbbm{P}_e$ is again ordered. Note the different time scales in (a), (b), and (c). 
		} \label{Fig::NPhotonBandwidthComp} 
	\end{figure*}

	Finally we illustrate the excitation probability dynamics.  Figure (\ref{Fig::NPhotonBandwidthComp}) shows $\mathbbm{P}_e$ for bandwidths $\Omega/\Gamma \in \{ 50 , 1, 1/20  \}$, chosen to illustrate three types of behavior.  In each subplot (a)-(c), excitation curves are plotted for photon numbers $N \in \{1,\dots,10\}$.  

	In \frf{Fig::NPhotonBandwidthComp}(a) a short pulse quickly excites the atom, which then decays into vacuum with rate $\Gamma$ after the wave packet leaves the interaction region.  Larger photon number corresponds directly to larger maximum excitation. In the intermediate bandwidth regime, $\Omega/\Gamma \approx 1$, excitations can be coherently exchanged between the atom and field, leading to oscillations in the excitation probabilities.  This continues until the wave packet leaves the interaction region as shown in \frf{Fig::NPhotonBandwidthComp}(b).  Similar damped Rabi oscillations were observed for large-photon-number coherent state wave packets in Ref.~\cite[Fig. 5]{WanSheSca10}.   For a single photon in the field, these oscillations are never seen due to the tradeoff between spectral bandwidth and photon density \cite{DomHorRit02, SilDeu03}. At the chosen bandwidth $\Omega/\Gamma = 1$, a single photon achieves the highest maximum excitation with maximum excitation falling off roughly with photon number in agreement with \frf{Fig::NPhotonScaling}.  Finally, in \frf{Fig::NPhotonBandwidthComp}(c) we see that an atom interacting with a long wave packet is excited and then decays well within the wave packet envelope and the $\mathbb{P}_e(t)$ curves are nearly symmetric around the peak of the wave packet for all photon numbers $N = \{1,\dots,10 \}$.

%% FIGURE 4:  N-Photon Excitation %%	

		%% SUBSUBSECTION:  Strong Coupling %%
		\subsubsection{Strong coupling} \label{SEC::StrongCoupling}

		%% FIGURE POSSIBLE:  Matching square-wave Rabi oscillations %%

	\begin{figure}[b]
	\begin{center}
	\includegraphics[width=1\hsize]{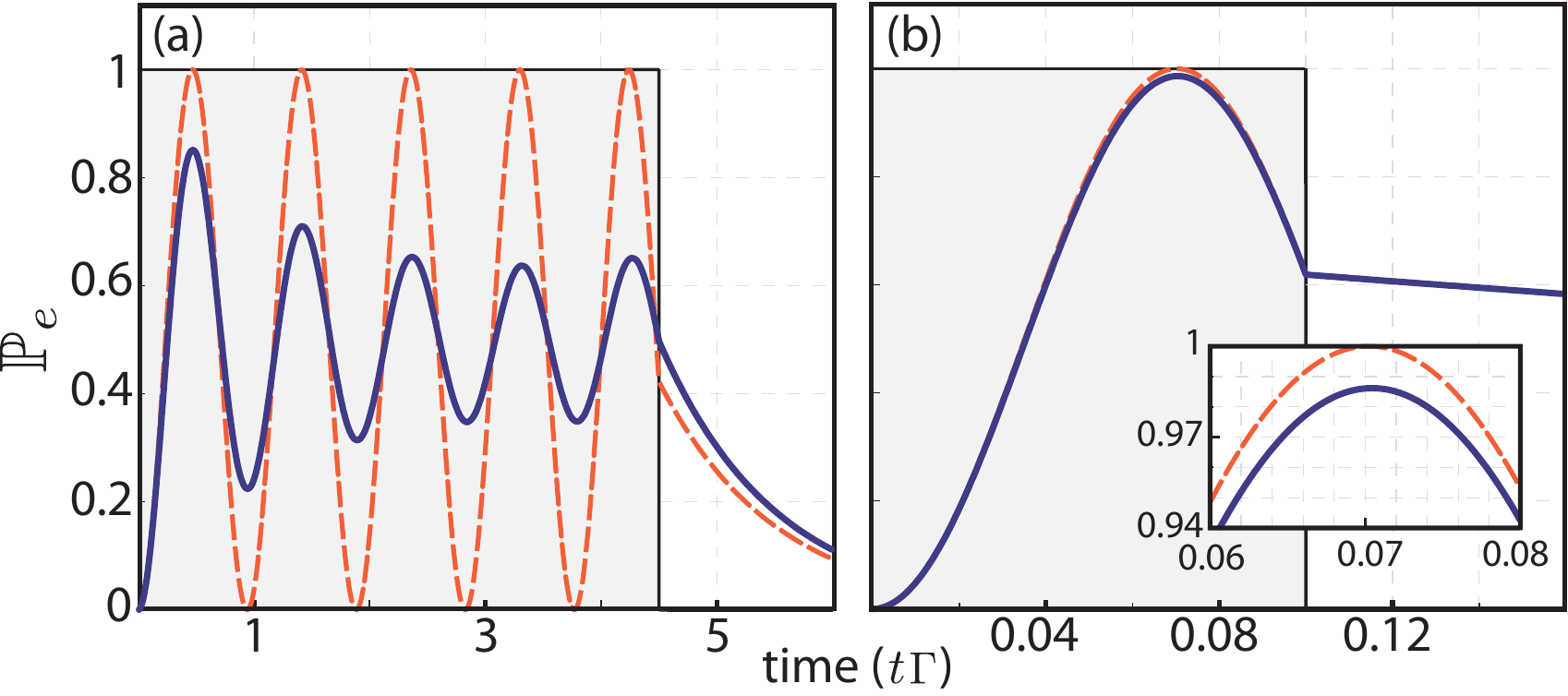}
		\caption{(Color online) Comparison of the numerically-calculated (dark blue) and analytically-predicted (dashed orange) Rabi oscillations for rectangular wave packets (normalized for clarity) with $N=50$ photons.  (a) Wave packet length $t_{\rm max}$ large compared to $1/\Gamma$. (b) Wave packet length approaching the limit $t_{\rm max} \ll 1/\Gamma$. We see increasing agreement between prediction and our numerics.} \label{Fig::IvanMatch} 
	\end{center}
	\end{figure}
	
	 The damped Rabi oscillations seen in \frf{Fig::NPhotonBandwidthComp}(b) suggest that there is a regime where coherent processes dominate over dissipation, known in cavity QED as the strong coupling regime. The authors of Ref.~\cite{SilDeu03} defined a strong coupling parameter (for {\em very} short rectangular wave packets): $\sqrt{N}g_{\rm eff} \gg \Gamma$ where $g_{\rm eff} = \xi(t) \sqrt{\Gamma_g}$. Specifically the wave packet was taken to be $\xi(t)= 1/\sqrt{t_{\rm max}}$ for times $t \le  t_{\rm max}\ll 1/\Gamma$ and zero otherwise. In this limit they showed that full Rabi oscillations for $N$ photons occur at frequency $\omega_R=g_{\rm eff}\sqrt{N}$.  In  \frf{Fig::IvanMatch} we compare their analytically-predicted excitation oscillations with our numerical calculations for $N=50$ photons.  In (a), the wave packet is long compared to $1/\Gamma$ and, while the oscillation frequencies match, the amplitudes do not due to dissipation.  For short wave packets, as seen in (b), coherent coupling prevails over dissipation, we see excellent agreement with the predicted frequency (in our parameters: $\omega_R=2 \xi(t)\sqrt{\Gamma_g N}$) and good agreement with the predicted amplitude.

	 For non-rectangular pulses the frequency of the Rabi oscillations is time-dependent as seen in \frf{Fig::NPhotonBandwidthComp}(b). We must account for the time variation of the wave packet $\xi(t)$ in order to define a more general strong coupling parameter.  To achieve strong coupling, the coherent coupling rate into the guided modes $\sqrt{N\Gamma_g}|\xi(t)|$ must dominate the total relaxation rate $\Gamma$.  We can immediately define the condition for instantaneous strong coupling:  $\sqrt{N\Gamma_g}|\xi(t)| / \Gamma \gg 1$.   However, in order to see interesting dynamics such as a complete Rabi oscillation, the coupling must remain strong over a characteristic timescale $\tau$. From this argument we define an {\em average} strong coupling parameter,
	\begin{align}\label{Eq::strongcoupling}
		\frac{\sqrt{N\Gamma_g}}{\Gamma \tau}\int_{t_s-\tau/2}^{t_s+\tau/2}dt  \,  |\xi(t)|\gg 1 \,\,\, \forall \, t_s.
	\end{align}
If, for any wave packet $\xi(t)$, there is a value of $t_s$ such that \erf{Eq::strongcoupling} is much greater than one, then average strong coupling has been achieved over the time window $\tau$.  

	A natural choice for $\tau$ is the characteristic decay time of the atom, $1/\Gamma$.  In \frf{Fig::StrongCoupling}(a) we present a contour plot of the average strong coupling parameter for Gaussian wave packets prepared in a single-photon Fock state ($N=1$).  Ideal coupling to the guided mode is assumed, $\Gamma_g = \Gamma = 1$.  We see that, for any bandwidth, maximum coupling occurs when the time window is centered at the Gaussian peak (indicated by the vertical, dashed white line) and that the strongest coupling is achieved for $\Omega/\Gamma = 4.$  Note that although the average strong coupling parameter for a single photon never exceeds one, for larger photon numbers the $\sqrt{N}$ factor can lead to significant coupling. In \frf{Fig::StrongCoupling}(b) the excitation probability dynamics are shown for an optimal bandwidth $\Omega/\Gamma = 4$ wave packet.   We see the appearance of damped Rabi oscillations when the wave packet has $N=50$ photons that are completely absent when only a single photon is in the field.  For comparison, a wave packet of bandwidth $\Omega/\Gamma = 2$ is shown in \frf{Fig::StrongCoupling}(c).  Even at this bandwidth, damped Rabi oscillations appear for $N=50$ photons, albeit with reduced contrast and frequency. 
 
%% FIGURE POSSIBLE:  Strong Coupling %%
	\begin{figure}[h!]
	\begin{center}
	\includegraphics[width=1\hsize]{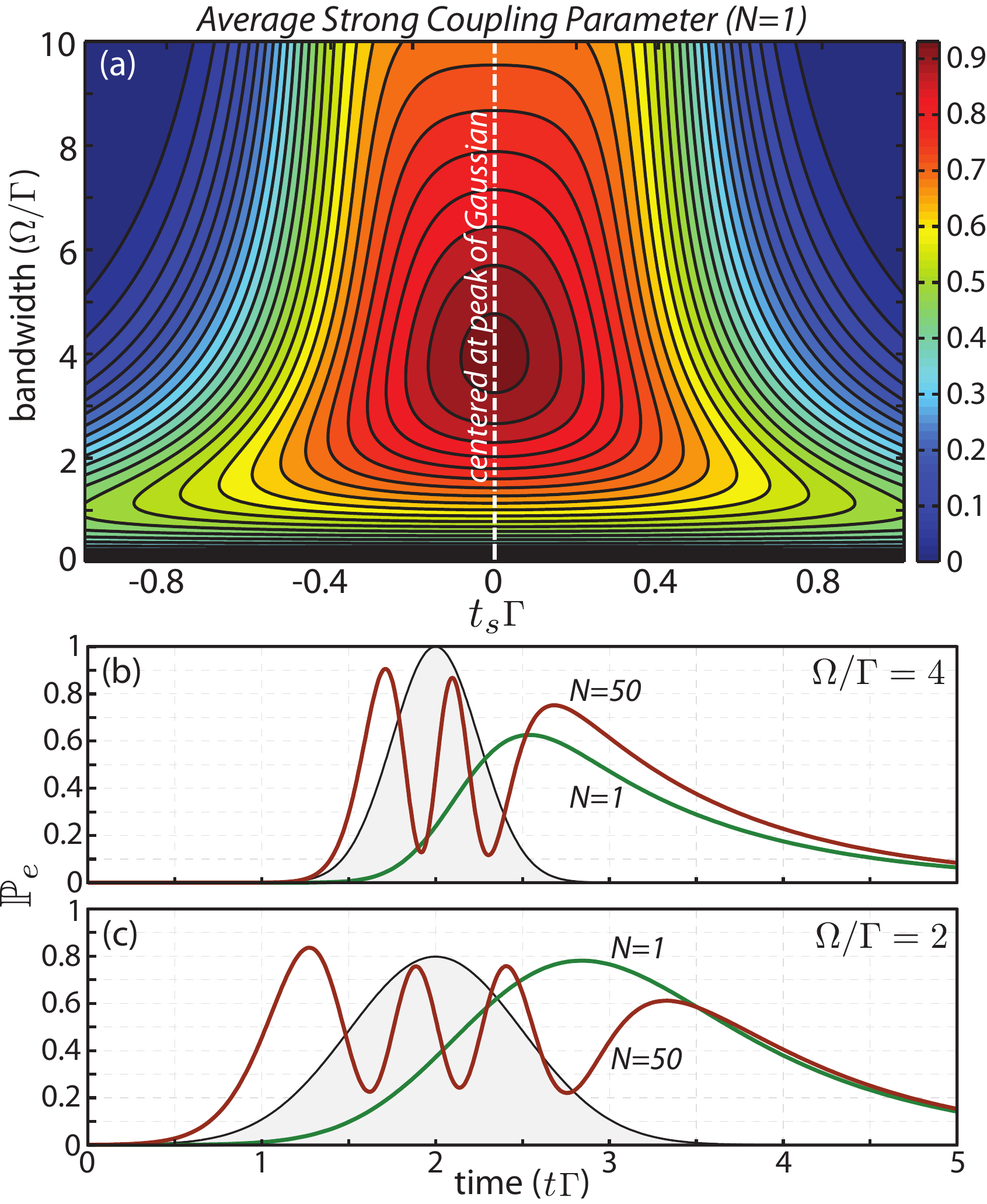}
		\caption{(Color online) (a) Contour plot of the average strong coupling parameter for a Gaussian wave packet prepared with a single-photon as a function of center of the time window ($t_s$) and bandwidth $\Omega/\Gamma$ (where $\tau =1/\Gamma$).  (b) and (c): Excitation probability of a two-level atom interacting with a wave packet of bandwidths $\Omega/\Gamma = 4$ for (b) and $\Omega/\Gamma = 2$ for (c). Only $N=1$ and $N=50$ photons are shown. The normalized wave packets $|\xi(t)|^2$ are shown in black filled grey.} \label{Fig::StrongCoupling} 
	\end{center}
	\end{figure}

%% SECTION:  Two mode Master Equations %%
\section{Two-mode Fock state master equations } \label{sec_2modefock}

	In this section we derive the master equations for a system interacting with an arbitrary combination of continuous-mode Fock states in two modes (spatial or polarization).  This generalization allows one to consider wave packets scattering off of atoms or addressing multiple dipole transitions, for instance.  The analysis for two modes is conceptually identical to but algebraically more complicated than the single-mode case.

	%% SUBSUBSECTION:  Two-mode stochastic propagator and quantum flow %%
	\subsection{Multi-mode It\={o} Langevin equations }

	The evolution of a system operator driven by multiple quantum noises is given by the multi-mode It\={o} Langevin equation, 
	\begin{align}\label{2mode_flow}
		dX & = \Big( i[ H, X] + \sum_i\mathcal{L}\dg[ L_i] X \Big)dt  + [ L\dg_i, X ] S_{ij} dB_j \nonumber \\ 
		& + S_{ij}\dg[ X, L_i] dB_j^{\dagger}  + ( S\dg_{ki} X S_{kj} - \delta_{ij} X) d\Lambda_{ij}. 
	\end{align}
where the modes are labeled by the subscripts $\{i,j,k\}$ and repeated indices are summed. $H$ is an external system Hamiltonian, the operator $L_{i}$ couples the system to the $i$th field mode, and the scattering operator $S_{ij}$ is constrained by: $S_{ik}S_{jk}\dg= \delta_{ij} I$ and $S_{ki}\dg S_{kj} = \delta_{ij} I$ (see \cite[
 Appendix A]{GouGohYan08}, \cite[Sec. IV]{GouJam09} and \cite{GouJam09b} and the references therein for more details on multi-mode QSDEs). Note that the subscript $t$ on the multi-mode quantum noise increments has been dropped for notational compactness in favor of the mode labels $\{ i,j \} $. The multi-mode quantum noise increments are defined,
	\begin{align}\label{2mode_noise}
		dB_i=\int_t^{t + dt} ds  \, b_i(s)  ,\,\, {\rm and}\,\, d\Lambda_{ij}=\int_t^{t + dt} ds  \, b_i^\dag(s) b_j(s).
	\end{align}
The composition rules for these quantum noises increments under Fock state expectation are
	\begin{align} \label{Eq::TwoModeItoTable}
	\begin{array}{c}
		dB_i dB\dg_j = \delta_{ij} dt,\,\,\, dB_i d\Lambda_{jk} = \delta_{ij} dB_k, \\
		d\Lambda_{ij} d\Lambda_{kl} = \delta_{jk} d\Lambda_{il},\,\,\, d\Lambda_{ij} dB\dg_k = \delta_{jk} dB\dg_i.
	\end{array}
	\end{align}

		%% SUBSUBSECTION:  Two-mode fock states
		\subsection{Two-mode Fock states }\label{2msec_fockstate}

	We consider the case where photons in mode one are prepared in a temporal wave packet $\xi(t)$ and those in mode two are in the wave packet $\eta(t)$.  The two-mode Fock state with $N$ photons in mode one and $Q$ photons in mode two is,
	\begin{align}
		\nn &\ket{N_\xi} \otimes \ket{ Q_\eta} =\frac{1}{\sqrt{N! Q!}} \left [ B_1\dg(\xi)\right ]^{N} \left [ B_2\dg(\eta)\right ]^{Q}\ket{0;0},
	\end{align}	
where the operators $B_i\dg(\cdot)$ are defined in \erf{eqBdef}.

	%% SUBSECTION:  General Two-Mode Master Equation %%
	\subsection{Two-mode Fock-state master equations for the system} \label{2msec_fockgen}

	Here we specialize the multi-mode equations, \erf{2mode_flow} and \erf{2mode_noise}, to two modes by restricting the indices to run over the mode labels $\{1,2 \}$. In Appendix \ref{Appendix::MultiMultimode} we show how do this calculation for any number of modes. We introduce notation for representing asymmetric expectations over two-mode Fock states,  
	\begin{align} \label{Eq::TwoModeGeneralDensityOps}
		\mathbbm{E}_{m,n;p,q}[ X(t) ]\! & =\! \Tr_{\rm sys + field} \!\left[ \!\left( \rho_{\rm sys}\! \otimes \!\op{m_\xi}{n_\xi} \!\otimes\! \op{ p_\eta}{ q_\eta } \right)\dg\!\! X(t)\! \right] \nonumber  \\
		& \equiv \!\Tr_{\rm sys} \left[ \varrho_{m,n;p,q}\dg(t)  X \right] ,
	\end{align}
which also defines the two-mode generalized density operators $\varrho_{m,n;p,q}$ in analogy with \erf{Eq::GeneralDensityOps}.  The reference field state is written as a tensor product where the labels $\{m,n\}$ refer to mode one and $\{p,q\}$ to mode two.  The two-mode Heisenberg master equations are found by taking field expectations over the equation of motion \erf{2mode_flow}. Thus, the action of the quantum noises on two mode Fock states is needed:
	\begin{subequations}
	\begin{align}
		dB_1\ket{n_\xi; q_\eta} & = \,dt\sqrt{n} \xi(t) \ket{n-1_\xi; q_\eta},\\
		dB_2\ket{n_\xi; q_\eta} & = \,dt\sqrt{q} \eta(t) \ket{n_\xi; q-1_\eta},\\
		d\Lambda_{11}\ket{n_\xi; q_\eta} & = dB_1\dg \sqrt{n} \xi(t) \ket{n-1_\xi; q_\eta},\\
		d\Lambda_{12}\ket{n_\xi; q_\eta} & = dB_1\dg \sqrt{q} \eta(t) \ket{n_\xi; q-1_\eta}.
	\end{align}
	\end{subequations}
The actions of $d\Lambda_{21}$ and $d\Lambda_{22}$ are similar.

	We then obtain the \sch-picture master equations with \erf{Eq::TwoModeGeneralDensityOps} and the cyclic property of the trace,
	\begin{widetext}
	\begin{align} \label{Eq::TwoModeME} 
		 \frac{d}{dt}& \varrho_{m,n;p,q}(t)  = - i[H,\varrho_{m,n;p,q}] + \big( \mathcal{L}[ L_1] + \mathcal{L}[ L_2] \big) \varrho_{m,n;p,q} \\ 
                  &  \nn+ \sqrt{m} \xi(t) [S_{i1} \varrho_{m-1,n;p,q}, L_i\dg  ] + \sqrt{p} \eta(t) [ S_{i2}\varrho_{m,n;p-1,q}, L_i\dg ] + \sqrt{n} \xi^*(t) [L_i, \varrho_{m,n-1;p,q} S_{i1}\dg  ] + \sqrt{q} \eta^*(t) [  L_i, \varrho_{m,n;p,q-1} S_{i2}\dg ] \\   
                  &\nn  + \sqrt{mn} |\xi(t)|^2 \left( S_{i1} \varrho_{m-1,n-1;p,q} S_{i1}^\dagger -  \varrho_{m-1,n-1;p,q}  \right)  
+ \sqrt{pq} |\eta(t)|^2  \left( S_{i2} \varrho_{m,n;p-1,q-1} S_{i2}^\dagger -  \varrho_{m,n;p-1,q-1}  \right) \\
                &\nn+ \sqrt{mq} \ \xi(t) \eta^*(t) \ S_{i1} \varrho_{m-1,n;p,q-1} S_{i2}^\dagger + \sqrt{np} \ \xi^*(t) \eta(t) \ S_{i2} \varrho_{m,n-1;p-1,q} S_{i1}^\dagger ,
	\end{align}
	\end{widetext}
where the subscript $i$ is summed over the mode labels. The initial conditions are
	\begin{align}
		\varrho_{m,n;p,q}(0) &=  \rho_{\rm sys} \quad & \mbox{if } m = n  \mbox{ and }   p=q \\
		\varrho_{m,n;p,q}(0) &=  0 \quad & \mbox{  if }  m \neq n  \mbox{  or } p\neq q. 
	\end{align}

	To solve a two-mode master equation with $N$ photons in mode one and $Q$ photons in mode two, $\rho_{\rm field} = \op{N_\xi}{N_\xi} \otimes \op{Q_\eta}{Q_\eta} $, we need to propagate $(N+1)^2\times(Q+1)^2$ coupled equations. As in the single-mode case the symmetries in the generalized density operators, $\varrho_{n,m; q,p} = \varrho_{m,n; p,q}\dg$, reduce the number of independent equations to $ \smallfrac{1}{4}(N+1)(N+2)(Q+1)(Q+2)$.

	%% SUBSECTION: General Input States %%
	\subsection{General input field states in the same wave packet }\label{sec_2modefockinput}

	So far we have only considered the case where the input fields in mode one and two are in ``pure'' Fock states, although we allowed for different wave packets. These results can be generalized to field states described by an arbitrary combination (superposition and/or mixture) of Fock states. Consider the state
	\begin{align} \label{Eq::2ModeGeneralField}
		\rho_{\mathrm{field}}&= \sum_{m,n,p,q=0}^\infty c_{m,n;p,q}  \op{n_\xi}{m_\xi} \otimes \op{ q_\eta}{ p_\eta }\\
		&= \sum_{m,n,p,q=0}^\infty c_{m,n;p,q}  \op{n_\xi ; q_\eta}{m_\xi; p_\eta }.
	\end{align}
As before, the coefficients, $c_{m,n;p,q}$, are constrained by the requirements of valid quantum states. For example the entangled N00N state for one photon is given by $\rho_{\mathrm{field}} = \half ( \op{1_\xi;0}{1_\xi;0}+\op{1_\xi;0}{0;1_\xi}+\op{0;1_\xi}{1_\xi;0}+\op{0;1_\xi}{0;1_\xi})$.

	When the input field is described by \erf{Eq::2ModeGeneralField}, the total system state is given by 
	\begin{equation}\label{Eq::gen_me2Mode}
		\varrho_{\rm total} (t)=\sum_{m,n,p,q}c_{m,n;p,q}^* \varrho_{m,n;p,q}(t),
	\end{equation}
where $\varrho _{m,n;p,q}(t)$ are the solutions to the master equations in \erf{Eq::TwoModeME}. The composition for expectation values is given by 
	\begin{equation}\label{Eq::EXPgen_me2Mode}
		\mathbb{E}_{\rm total}[X(t)] =\sum_{m,n,p,q}c_{m,n;p,q}\mathbb{E}_{m,n;p,q}[X(t)].
	\end{equation}
As before, the conjugate coefficients in \erf{Eq::gen_me2Mode} come from the Hilber-Schmidt inner product, \erf{Eq::HSInnerP}.  This technique also applies to the output field quantities in \srf{2msec_fockgen_output}.

	%% SUBSECTION:  Two-mode Output Field Master Eqs %%
	\subsection{Two-mode output field quantities}\label{2msec_fockgen_output}

	The output field equations for two modes are significantly more complicated than the single-mode case because one can consider linear combinations of the modes. Thus, there is a continuum of possible of output photon fluxes and field quadratures. Here we focus on photon flux and field quadrature observables that are diagonal in the modes.  More complicated output observables that combine both modes can be obtained using beam splitter relations -- effectively, a change of basis -- as described in Ref.~\cite{GouJam09}.

		%% SUBSUBSECTION:  Two-mode output intensity
		\subsubsection{ Photon flux}

The number of photons scattered from mode $j$ into mode $i$ in the interval $t$ to $t+d t$ is given by $d\Lambda_{ij}^{\rm out}$.  Its equation of motion is
	\begin{align} \label{Eq::dLambdaTwoMode}
	d \Lambda_{ij}^{\rm out} & =  L_i^\dagger L_j dt + L_i^\dagger S_{jk} dB_k
+ S_{ik}^{ \dagger} L_j dB_k^{ \dagger} +S_{ki}^\dagger S_{lj} d\Lambda_{ij} .
	\end{align}
Any possible two-mode photon counting distribution is given by taking expectations of \erf{Eq::dLambdaTwoMode}. For example, tracing over the system and field for $d\Lambda_{11}$ gives the mean photon flux in mode one,
	\begin{align} \label{Eq::TwoModeFieldME_lambda}
		\frac{d}{dt} & \mathbb{E}_{m,n;p,q} [\Lambda_{11}^{\rm out}(t)] = \mathbb{E}_{m,n;p,q}[{L}_1\dg L_1] \\
		& + \sqrt{m} \xi^*(t) \mathbb{E}_{m-1,n;p,q} [ S_{11}\dg L_{1} ]  \nn \\
		& + \sqrt{p} \eta^*(t) \mathbb{E}_{m,n;p-1,q} [S_{12}\dg, L_1 ] \nonumber \\
		& + \sqrt{n} \xi(t) \mathbb{E}_{m,n-1;p,q}[ L_1\dg S_{11} ] \nonumber \\
		& + \sqrt{q} \eta(t) \mathbb{E}_{m,n;p,q-1} [ L_1\dg {S}_{12} ], \nonumber \\
		& + \sqrt{mn} |\xi(t)|^2 \sum_{i,j}  \mathbb{E}_{m-1,n-1; p,q}[{S}\dg_{i1} {S}_{j1} ]. \nonumber 
	\end{align}
The equation for mode two follows similarly.

			%% SUBSUBSECTION:  Two-mode output quadratures	
			\subsubsection{ Field quadratures }

	The output quantum noise in mode $i$ is given by
	\begin{align}
 		d B_i^{\rm out} = S_{ij} dB_j + L_i dt.
	\end{align}	
	Just as in the single-mode case, field quadratures are Hermitian combinations of $B_i$ and $B_i\phantom{}^\dagger$.  For instance, the field quadrature in mode one, $Z_1^{} = e^{i\phi} B_1^{} + e^{-i\phi} B_1^{}\!\!\phantom{}^\dagger$.  The equation of motion for the mean ouput field quadrature $Z_1^{\rm out}$, or homodyne current, after the interaction is, 
	\begin{align} \label{Eq::TwoModeFieldME}
		\frac{d}{dt}   \mathbb{E}_{m,n;p,q} [Z_1^{\rm out}(t)] =& \mathbb{E}_{m,n;p,q}[e^{i\phi} L_1+ e^{-i\phi} L_1\dg] \\
		&  + e^{i \phi} \sqrt{m} \xi^*(t) \ \mathbb{E}_{m-1,n; p,q}[S_{11}\dg  ]\nonumber\\
		&  + e^{i \phi} \sqrt{p} \eta^*(t)\ \mathbb{E}_{m,n; p-1,q}[ S_{12}\dg ] \nonumber \\
         		&  + e^{-i \phi} \sqrt{n} \xi(t) \ \mathbb{E}_{m,n-1; p,q}[S_{11} ]\nonumber\\
		&  + e^{-i \phi} \sqrt{q} \eta(t)\ \mathbb{E}_{m,n; p,q-1}[ S_{12}].\nonumber 
	\end{align}
The equations for $Z_2^{\rm out}$ follow similarly.

	%% SUBSECTION:  General Two Mode States %%
	\subsection{General two-mode $N$-photon states}

	The formalism developed in \srf{Sec::GeneralWavepackets} suffices to describe arbitrary states in each mode separately and thus is directly applicable to the two-mode master equations. 

	A slightly more general case is when there are $m$ photons in mode one and $N-m$ photons in mode two with an arbitrary spectral distribution function (such two-mode states can be entangled in the spectral degree of freedom).  These states can be written
	\begin{align}\label{eq:blahb}
		\nn \ket{\psi_N}=& \int d\omega_1\dots d\omega_{N}  \, \tilde\xi_{N}(\omega_1,\dots,\omega_{N}) \\
		&  \quad\quad \times  b_1^{\dagger}(\omega_1)\dots b_1^{\dagger}(\omega_{m})b_2^{\dagger}(\omega_{{m}+1})\dots b_2^{\dagger}(\omega_{N})\ket{0}.
	\end{align}
With a straightforward generalization of the formalism developed in Appendix (\ref{Appendix::multiphotonStates}) and \srf{Sec::GeneralWavepackets} one can derive master equations for states of the form of \erf{eq:blahb}. 

	Even more general is an $N$-photon state distributed over two modes $b_1$ and $b_2$,
	\begin{align}\label{Eq::SuperGen2mode}
		\nn \ket{\psi_{N}}=& \int d\omega_1\dots d\omega_N\, \tilde\xi_{N}(\omega_1,\dots,\omega_N)\\
		&  \quad\quad\quad\quad \times \prod_i^N \big (\alpha_i b_1^{\dagger}(\omega_i)+\beta_i b_2^{\dagger}(\omega_i)\big)\ket{0}
	\end{align}
where $\alpha_i$ and $\beta_i$ are weights for modes one and two, respectively. For example, if we set all the $\alpha_i=0$ in \erf{Eq::SuperGen2mode} then there would be $N$ photons in mode two.  For a small number of photons it is tedious, but possible, to write down the occupation number representation of the state in \erf{Eq::SuperGen2mode}. Finding an efficient representation for such state with arbitrary $N$ is an open problem and would allow a derivation of general two-mode master equations.

%% SECTION:  Two mode scattering off a two level atom %% 
\section{Two-mode example: Fock-state scattering from a two-level atom} \label{Sec::2modeExample}

	In this section we illustrate the use of our two-mode formalism by examining the photon flux of the transmitted and reflected fields when {Fock states} are incident on a two-level atom \cite{DroHavBuz00, DomHorRit02, CheWubMor11,ZheGauBar10,SheFan07b,ZhoGonSun08,LonSchBus09,SheFan05,Roy10}.  The two modes are the forward- and backward-propagating fields, as in a tightly-confined waveguide QED setting \cite{SheFan07b,ZheGauBar10}.  As before we specialize to a Gaussian wave packet $\xi(t)$ described by \erf{Eq::gau_xi}. The master equation parameters we use are again those for dipole coupling without external Hamiltonian drive: $H=0$, $L_i= \sqrt{\Gamma_i} \op{g}{e}$, $S_{ii}=I$, $S_{ij}=0$ for $i\neq j$, and the coupling rate is chosen to be $\Gamma_i = 1/2$.  The forward-propagating field is prepared in a Fock state with $N \in \{1,\dots,5\}$ photons while the backward mode is initially in vacuum; that is, $\ket{\psi_{\rm field}} = \ket{N_\xi; 0}$.

%% FIGURE 5:  Wavepacket Reflection %%

	\begin{figure}[b]
	\begin{center}
	\includegraphics[width=1\hsize]{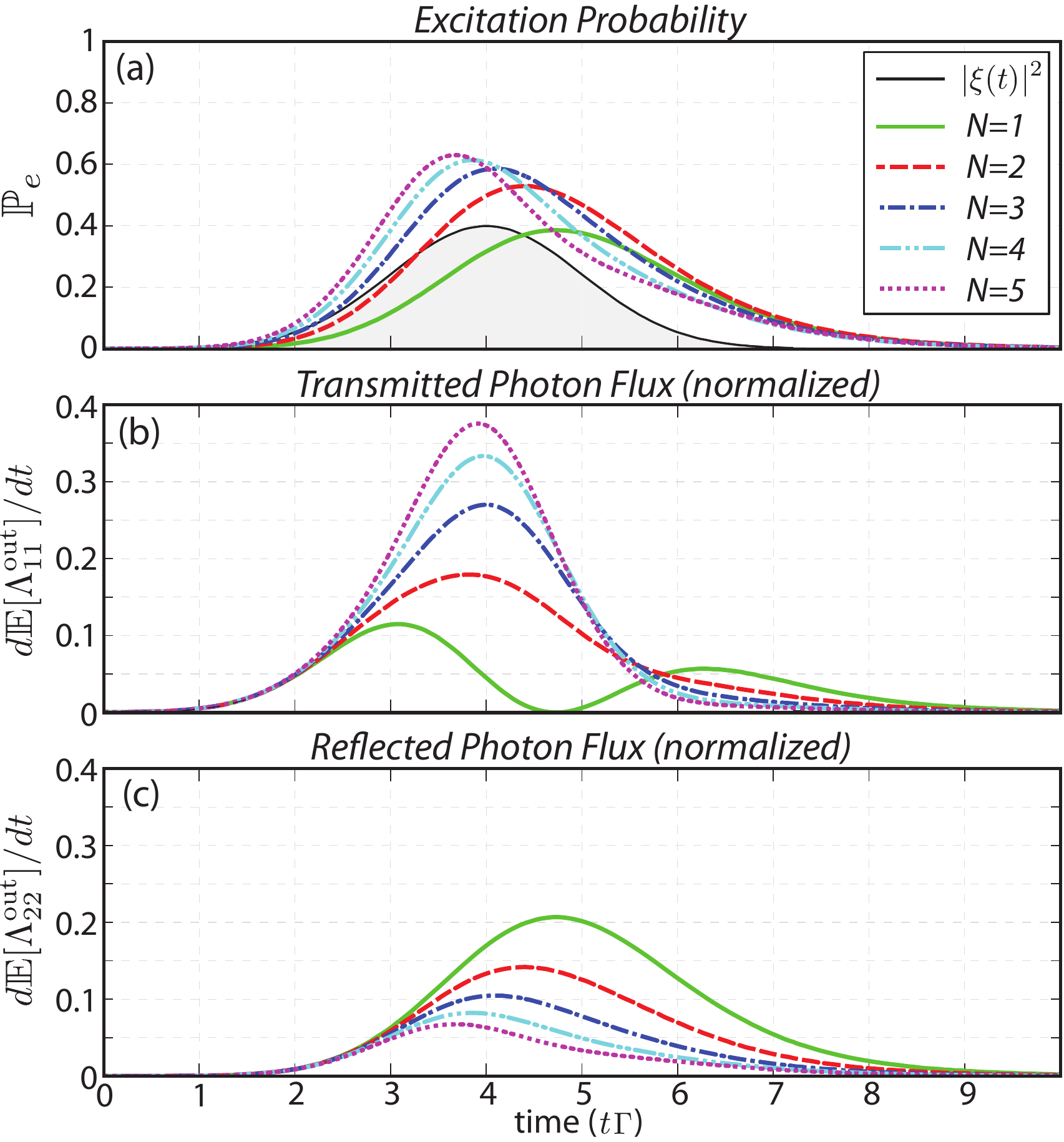}
		\caption{(Color online) Scattering of a Gaussian wave packet of bandwidth $\Omega/\Gamma=1$ from a two-level atom.  The wave packet $|\xi(t)|^2$ (black filled grey) is prepared with $N \in \{1,\dots,5 \}$ photons. (a) Excitation probability.  Photon flux of the transmitted (b) and reflected (c) fields, normalized to input photon number.} \label{Fig::2m_reflect}
	\end{center}
	\end{figure}

 In \frf{Fig::2m_reflect}(a) we plot the excitation probability $\mathbb{P}_e$ for a two-level atom interacting with a wave packet with bandwidth $\Omega/\Gamma=1$. The photon flux of the transmitted and reflected fields is plotted in Figures \ref{Fig::2m_reflect}(b) and (c), normalized to the number of input photons $N$. 
	
	We first examine the single-photon input state (solid green curves). While absorbing the photon, the atom has a substantial $\mathbb{P}_e$. The two peaks in the transmitted flux correspond to the attenuated input wave packet and the contribution from remission into the forward mode~\cite{DroHavBuz00}.  Notice the dip between the peaks occurs when there is a large atomic excitation. Consequently this dip in the transmitted photon flux is due to atomic absorption and destructive interference with the incoming wave packet \cite{DroHavBuz00,LonSchBus09,CheWubMor11,SheFan05}. Conversely, energy from the field that is not absorbed is scattered into the backward mode through the reemission process~\cite{DroHavBuz00}.  For $N>1$, we see that the excitation probability is comparable to that for a single photon, but the relative transmitted and reflected photon fluxes are quite different. In particular the ratio of transmitted to reflected flux increases with $N$.

	In order to understand this phenomena it is necessary to consider the normalized transmitted and reflected photon numbers in the long-time limit ($\mathbbm{E}[\Lambda_{11}]$ and $\mathbbm{E}[\Lambda_{22}]$) at different bandwidths \cite{DroHavBuz00,ZheGauBar10}. In \frf{Fig::poindexter} we explore this issue numerically. Recall that the reflection process is facilitated by absorption and then reemission into the backward mode. Thus one would expect reflection to dominate for small bandwidth wave packets, which is indeed what is seen in the left hand side of \frf{Fig::poindexter}. In the large bandwidth limit very little of the wave packet is near resonance with the atomic transition so no absorption occurs and the wave packet is transmitted. The bump in the $N>1$ transmission and reflection curves is a consequence of an effective photon-photon interaction \cite{DeuChi92, SheFan07b,ZheGauBar10}.  By calculating the scattering eigenstates, Zheng et al. found ``multi-photon bound states"~\cite{ZheGauBar10} which can increase transmission in that bandwidth region. 
	
	It is also possible to examine scattering between the forward and backward modes, as was studied in Ref.~\cite{ZheGauBar10}, by propagating the equations for $\Lambda_{12}$ and $\Lambda_{21}$; however, we omit this analysis for brevity.

%% FIGURE 6:  3 Photon Transmission and Reflectance %%

	\begin{figure}[t]
	\begin{center}
	\includegraphics[width=1\hsize]{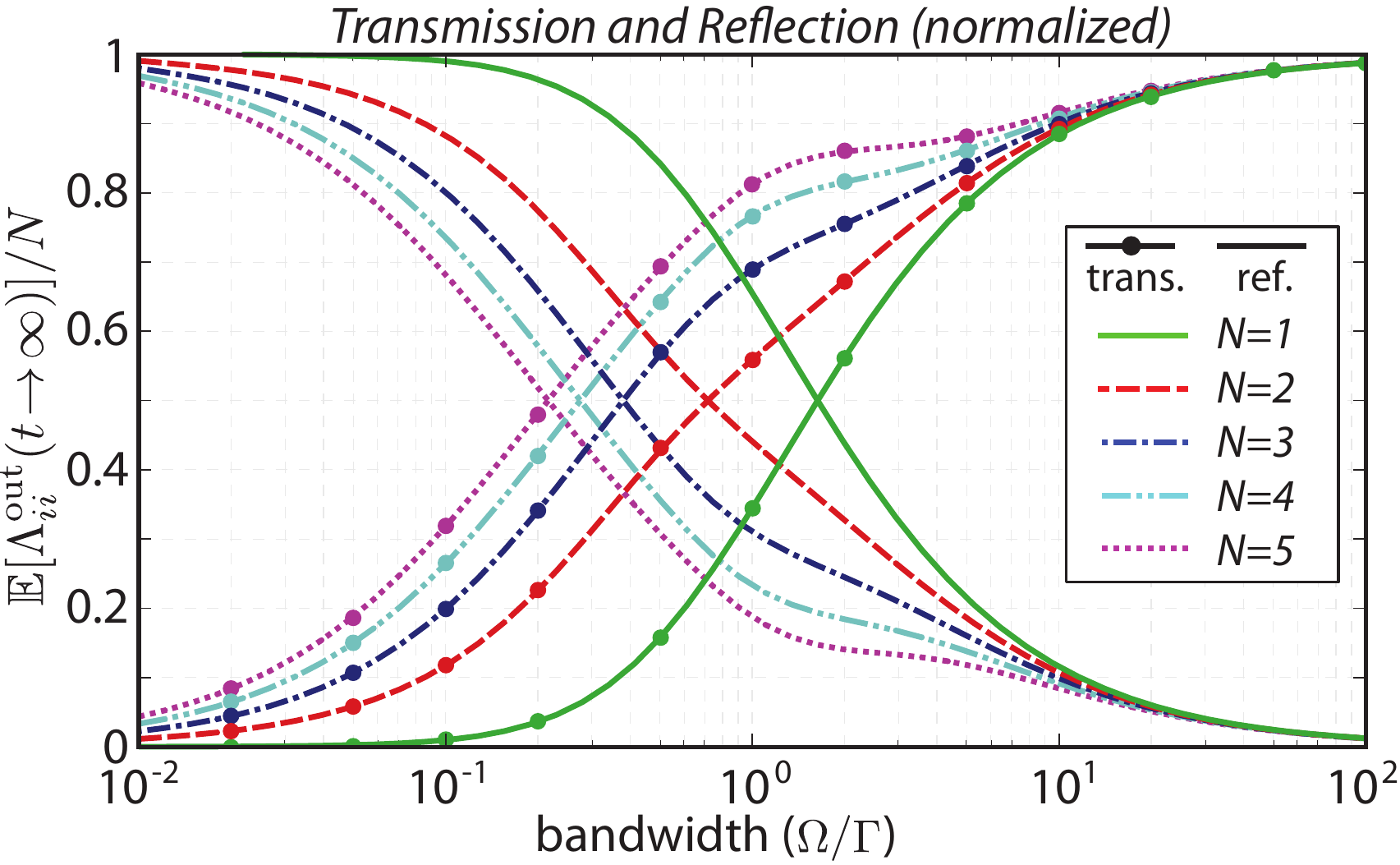}
		\caption{(Color online) Normalized transmission and reflection for Gaussian wave packets,  prepared with $N\in \{1,\dots,5\}$ photons, with bandwidths $\Omega/\Gamma$ scattering from a two-level atom.  The left (right) side represents long (short) temporal wave packets. For larger photon number, note the increased transmission at intermediate bandwidths. } \label{Fig::poindexter}
	\end{center}
	\end{figure}

%%% SECTION:  Discussion and Conclusion %%%
\section{Discussion}\label{sec_dis}

	In this paper we have derived master equations for an arbitrary quantum system interacting with a continuous-mode Fock state in one or two modes (spatial and/or polarization).  We generalized these results to include superpositions and mixtures of $N$-photon states with arbitrary spectral distribution functions, and thus we can describe interaction with very general states of light. 

	The power of our formalism lies in its direct applicability to more general systems of interest in quantum optics such as multi-level atoms, symmetrically-coupled atomic ensembles, and continuous variable systems such as nano mechanical resonators. For example, it is possible to reproduce the cavity-mediated, single-photon pulse shaping results of Ref.~\cite{Mil08}. First we identify that $ H=0$,  $L= \sqrt{\gamma} a$, and $S=I$ are the relevant substititutions. Then, our expression for the output photon flux, \erf{Eq::SingleModeFieldME_lambda}, is equivalent to Equation (22) in Ref.~\cite{Mil08} for one photon (i.e. in our equations set $N_{\rm max}=1$) after some algebraic gymnastics. 

	As pedagogical examples, we studied features of Fock states interacting with a two-level atom in one and two modes. In the single-mode model [\srf{Sec::2LevelExcitation}] we saw the maximum excitation probability $\mathbb{P}_e^{\rm max}$ was low for both small ($\Omega / \Gamma \ll10^{-1}$) and large ($\Omega / \Gamma \gg10^{2}$) bandwidths. The low $\mathbb{P}_e^{\rm max}$  for small bandwidths, centered at the atomic resonance, might seem counter intuitive. In the time domain the corresponding wave packet is broad, nevertheless the near-resonant photons all get absorbed, but are immediately reemitted by the vacuum coupling, which leads to a small average $\mathbbm{P}_e$. This intuition is confirmed in the two-mode simulations, presented in \srf{Sec::2modeExample}, where wave packets with small bandwidths are nearly perfectly reflected. The reflection is mediated by photon absorption and the consequent reemission, which is directionally unbiased.  However, destructive interference between the incoming wave packet and the transmitted mode results in reflection only; i.e. the atom can act as a perfect reflector. 
	
	A detailed investigation of this phenomenon requires access to the individual quantum trajectories \cite{CarBook93} rather than the ensemble averaged evolution given by the master equations. For a single photon, a step towards the differential equations for the quantum trajectories, known as stochastic master equations or Òquantum filtersÓ  \cite{BouvanHJam07}, was given in Ref. \cite{GheEllPelZol99}. Gheri et al. \cite[Sec. V]{GheEllPelZol99} suggested using the cascaded systems approach \cite{Gar93,Car93} to determine the conditional evolution of a single photon interacting with a quantum system. This suggestion has become a standard approach see e.g. Ref.~\cite{BasAkrMil12}.  However an elegant alternative exists. Recently the single-photon quantum filtering equations were derived from first principles for homodyne \cite{GJNphoton,GJNCgen} and photon-counting \cite{GJNCgen} measurements of the output fields.  We are presently extending these to Fock states in one and two modes. Access to the conditional states would allow for measurement-based feedback control \cite{WisMilBook}.

	A number of interesting applications of our formalism remain to be explored, including the investigation of pulse shaping for few-photon states, high efficiency quantum memories, and mediated photon-photon interactions.  Our formalism is particularly applicable to quantum networks \cite{Kimble08, MoeMauOlm07}. Recently, the theory of cascaded quantum systems \cite{Gar93, Car93} has been formalized to the point where simple rules for composing modular quantum optical systems into a network have been developed \cite{YanKim03a,YanKim03b,GouJam09,GouJam09b,GouJam10}. One needs only the $(S,L,H)$-tuple of each module specified in order to perform network analysis and simplification. As our description of the system, input, and output fields is also in terms of a $(S,L,H)$-tuple, it is likely that our formalism can be ported to this setting. 

{\em Acknowledgements:} The authors acknowledge helpful discussions with Yimin Wang, Ji\v{r}\'{\i} Min\'a\ifmmode \check{r}\else \v{r}\fi{}, Valerio Scarani, Norman Yao, Brad Chase, and Ivan Deutsch. In particular the authors would like to thank Carl Caves for carefully reading part of this manuscript and making many helpful suggestions.  JC would also like to thank the participants in the experimental session of the CQT Workshop on Quantum Tomography for valuable comments as well as Matt James, Hendra Nurdin, John Gough, and Masahiro Yanagisawa for many discussions about QSDEs. BQB would like to thank Valerio Scarani for financial support during a research visit.  BQB, RLC, and JC acknowledge financial support from NSF Grant No. PHY-0969997, No. PHY-0903953, and No. PHY-1005540, ONR Grant No. N00014-11-1-008, and AFOSR Grant No. Y600242. AMB acknowledges support from DARPA (QuBE).

%%% APPENDICES %%%
\appendix

 %%% APPENDIX I: QUANTUM NOISE %%%
 \section{Quantum noise and quantum stochastic calculus}\label{quan_stoch_calc}

	 A rich mathematical machinery forms the foundation for the manipulation of QSDEs and their derivation from physical systems.  Here we only touch the surface commensurate with our purposes; an interested reader is directed to Refs.~\cite{YanKim03a, vanHStoMab05a, DumParZol92,HudPar84,Par92,Gou06,GarCol85,GarZol00,Bar06, Accardi, WisMilBook} for a more rigorous and detailed analysis.
	
	We present an introduction to the formalism of quantum stochastic calculus through the canonical example of a two-level atom interacting with a quantized, one-dimensional field.  The atomic raising and lowering operators are $\sigma_+ = \op{e}{g}$ and $\sigma_-= \op{g}{e}$ with transition frequency $\omega_0$.  The field is described by creation and annihilation operators, $a\dg(\omega)$ and $a(\omega)$, obeying the commutation relation $[ a(\omega), a\dg(\omega')]  = \delta(\omega -\omega')$.  The interaction-picture coupling between the atom and the field, within the rotating wave approximation, is
	\begin{equation}   \label{Eq::HdpRWAInt}
 		H_{\rm int}(t) =\,-i \hbar \sigma_+ \int d\omega \, \kappa(\omega) a(\omega) e^{- i (\omega - \omega_0) t } + \rm{H.c.},
	\end{equation}
where the dipole coupling, $ \kappa(\omega) = |\bra{e}d\ket{g}| \sqrt{\omega/4 \pi \epsilon_0 \hbar c A}$, has units of $\sqrt{ \rm frequency } $ and $A$ is the effective transverse cross-sectional area of the mode, see Domokos et al. \cite{DomHorRit02}.

	%%% SUBSECTION: The quantum white noise limit %
	\subsection{The quantum white noise limit}\label{SEC::QWNL}

To take the quantum white noise limit we first assume weak coupling, i.e. that $|\kappa(\omega)|^2 \ll \omega_0$.  When $\kappa(\omega)$ is slowly varying around $\omega_0$, we make the Markov approximation that the atom has a flat spectral response; i.e. $\kappa(\omega) \rightarrow \kappa(\omega_0)$.  This implies that the correlation time of the field is short compared to the slowly-varying interaction time, $\tau_s  \approx 1/|\kappa(\omega_0)|^2 $.  From the perspective of atomic operators, the field is $\delta$-correlated in time and retains no memory of its past interactions. In this limit we can introduce the following field operators,
	\begin{equation}
    		b(t) = \frac{1}{\sqrt{2\pi} } \int d\omega \, a(\omega) \, e^{-i (\omega - \omega_0) t},
	\end{equation}
which obey the commutation relation $[b(t),\, b^\dag(t')] = \delta(t - t')$. For classical stochastic processes, $\delta$-correlation implies white noise, so the operators $b(t)$ and $b^\dagger(t)$ are dubbed \emph{quantum white noise operators}.  Recast in terms of these operators the interaction Hamiltonian is
	\begin{equation} \label{Eq::WhiteNoiseDipHam}
 		H_{\rm int}(t) =\,i \sqrt{\gamma} \left( \sigma_-\, b^\dag(t) - \sigma_+\, b(t) \right)
	\end{equation}
where we define $\kappa(\omega_0) = \sqrt{\gamma/2\pi}$ and set $\hbar = 1$.  

Under the white noise-driven Hamiltonian in \erf{Eq::WhiteNoiseDipHam}, the system and the field undergo joint unitary evolution via the propagator {    $U(t)$} that satisfies the Schrodinger equation,
	\begin{equation}\label{Eq::USchEqn}
		\frac{d}{d t} U(t) =  \sqrt{\gamma} \left( \sigma_-\, b^\dag(t) - \sigma_+\, b(t) \right) U(t).
	\end{equation}
This expression defies rigorous mathematical definition due to the singular commutation relation of the operators $b(t)$ and $b^\dagger(t)$.  To remedy this we first consider the quantum stochastic processes,
	\begin{align}\label{Eq::wotupdawg}
		B_t = \int_0^t ds \, b(s) \quad \text{and} \quad B^\dagger_t = \int_0^t ds \, b^\dagger(s).
	\end{align}
The singular nature of the quantum white noise operators can be removed by expressing \erf{Eq::USchEqn}  in terms of the continuous differential increments $dB_t$ and $dB\dg_t$ of \erf{Eq::wotupdawg}:
	\begin{equation}
		\int_t^{t + d t}\! ds  \, b(s) \mapsto dB_t \,\,\, \text{and }\,\,\, \int_t^{t+ d t} \!ds  \, b^\dag(s)  \mapsto dB_t\dg.
	\end{equation} 
These are the quantum, non-commuting analogues of the classical Wiener process and are referred to generically as \emph{quantum noise increments}.  Now equation (\ref{Eq::USchEqn}) can be recast in differential form:
	\begin{equation}\label{Eq::UStratonovich}
    		dU_t =  \sqrt{\gamma} \left( \sigma_-\, dB^\dag_t - \sigma_+\, dB_t \right)\circ U_t.
	\end{equation}
Although technically an integral equation, this is referred to as a \emph{quantum stochastic differential equation} (QSDE).

	In contrast to ordinary differential equations, white noise QSDEs have equivalent but non-identical representations.  Equation (\ref{Eq::UStratonovich}) is an example of a Stratonovich QSDE, identified by the notation $dB_t \circ U_t$, which indicates the ordering of $dB_t$ and $U_t$ is important; i.e. that they do not commute.  Stratonovich QSDEs arise as the natural form for the quantum white noise limit of physical processes \cite{Gou06} and follow the rules of standard calculus.  More amenable for our purposes is the It\={o} form of a white noise QSDE.  The quantum It\={o} integral is defined such that the integrand and the operator differential, $dB_t$, act on independent time intervals and therefore commute, which is useful for taking expectations.  Thus, we will work exclusively with QSDEs in It\={o} form, denoted simply by $dB_t U_t$.  However, the It\={o} form brings the burden of its own calculus, which requires that differentials be taken to second order.

	Performing the conversion from Stratonovich to It\={o} form \cite{GarZol00,Gou06} on \erf{Eq::UStratonovich} and renormalizing a trivial energy shift, we obtain the QSDE for the unitary time-evolution operator
	\begin{equation}\label{Eq::UIto}
   		 dU_t =   \left( \sqrt{\gamma}\, \sigma_-\, dB^\dag_t - \sqrt{\gamma}\, \sigma_+\, dB_t  - \half \gamma 		\sigma_+ \sigma_- \, dt \right) U_t.
	\end{equation}
The first two terms represent the atomic dipole coupling to the quantum noise increments, and the third deterministic term is an artifact of the transformation from Stratonovich to It\={o} form, known as the \emph{It\={o} correction}.

	%%% SUBSECTION: General stochastic time evolution operator %%%

	\subsection{General stochastic time evolution operator} \label{SubSection::GeneralPropagator}

The quantum white noise limit can be extended to include coupling of a system operator $\tilde{S}$ to the number of photons in the field at time $t$.  This interaction Hamiltonian is
	\begin{equation} \label{Eq::HamNum}
		 H_{\rm num}(t) = \tilde{S} \  b^\dag(t) b(t).
	\end{equation}
From this Hamiltonian we identify a third fundamental quantum noise which can drive the system in the white noise limit,
	\begin{align}
		\Lambda_t = \int_0^t ds \, b\dg(s) b(s) 
	\end{align}
which has increments
	\begin{align}\label{Eq:dLambda_yo}
		\int_t^{t + d t} ds  \, b^\dag(s) b(s)  \mapsto d\Lambda_t.
	\end{align}
Including the possibility of an external system Hamiltonian $H$, the most general QSDE for the time evolution operator in one mode has the form \cite{GouJam09}
	\begin{align}   \label{Eq::QsdeUnitary}
 		dU_t = \Big\{ - (\half & L\dg L  + iH )dt  - L\dg  S dB_t   \\
 		\nn & + L dB\dg_t + (S - I )d\Lambda_t  \Big\} U_t.
	\end{align}
This equation describes the coupling of system operators $L$, $L\dg$, and $S$ to the quantum noises $dB\dg_t$, $dB_t$, $d\Lambda_t$, and $I$ is the identity operator.  The system operator $S$ can be found from the bare Hamiltonian coupling of $\tilde{S}$ in \erf{Eq::HamNum} with rules described in Ref. \cite{GoughvHan07}.

	%%% SUBSECTION: Ito Langevin Equations %%%
	
	\subsection{It\={o} Langevin equations} \label{App::ItoLangEQS}

	The time evolution operator in \erf{Eq::QsdeUnitary} allows us to calculate the equation of motion for an operator $O$.  Since we work with It\={o} QSDEs, this requires taking differentials to second order,
	\begin{align} \label{Eq::HEOMArb}
		d(U_t \dg O U_t) = d U_t\dg OU_t + U_t\dg O dU_t + d U_t\dg O d U_t.
	\end{align}
Note that in the literature one may encounter the ``quantum flow" notation where an operator $O$ at time $t$ is given in the Heisenberg picture by $j_{t}( O ) \equiv U_t \dg O U_t $. When manipulating QSDEs such as \erf{Eq::HEOMArb} one encounters products of the quantum noise increments.  Under vacuum expectation the rules for these products are given by the vacuum It\={o} table
	\begin{equation} \label{Eq::ItoTableAppx}
	\begin{tabular}{l |llll}
		$\times $        & $dB_t$ & $d\Lambda_t $ & $dB^{\dag }_t$ & $dt$ \\ \hline
		$dB_t$              & 0         & $dB_t$               & $dt$                  & $0$ \\
		$d\Lambda_t $ & 0        & $d\Lambda_t$  & $dB^{\dag }_t$  & $0$ \\
		$dB^{\dag }_t$ & 0         & 0                      & 0                       & 0\\
		$dt$                & 0         & 0                      & 0                       & 0 
	\end{tabular} ,
	\end{equation}
where we take the row and multiply by the column (row $\times$ column) to obtain the resulting product under vacuum.

	With \erf{Eq::HEOMArb} and \erf{Eq::ItoTableAppx} we can write down the It\={o} QSDE for an operator $X \otimes I_{\rm field}$,
	\begin{align} \label{Eq::dXAppendix}
		\nn dX =&( i[H, X] + \mathcal{L}\dg[L]X) dt   + [L\dg,X]S dB_{t} \\
		& +S\dg[X,L] dB\dg_{t}  + (S\dg XS-X) d\Lambda_{t},
	\end{align}
referred to as an \emph{It\={o} Langevin equation}. Further, we can write down the It\={o} Langevin equation for output field quantities, such as the quantum noise $B_t^{\rm out}$,
 	\begin{align}  \label{Eq::dBAppendix}
		dB_t^{\rm out} = L dt + S dB_t,
	\end{align}
and photon number $\Lambda_t^{\rm out}$,
	\begin{align}  \label{Eq::dLambdaAppendix}
		d\Lambda_t^{\rm out} &=   L\dg Ldt +  L\dg S dB_t + S\dg L dB\dg_t  + S\dg S d\Lambda_t,
	\end{align}
 where $S\dg S = I$.

 	%%% SUBSECTION:  Multi-mode time evolution operator %%%
	 \subsection{Multi-mode time evolution operator}
	 The evolution of a system driven by multiple quantum noises is given by the QSDE for the multi-mode time evolution operator, 
	\begin{align}  \label{a2mode_qsdeunitary}
		\nn dU_t =& \Big\{ ( S_{ij} - \delta_{ij} I)d\Lambda_{ij} - L_{i}^\dagger S_{ij} dB_j + L_i dB_i^\dagger \\
		& \,\,- (\half L_i^{\dagger} L_i +i H )dt  \Big\} U_t, 
	\end{align}
where $L_{i}$ is the coupling between the $i$th mode and the system, $H$ is an external Hamiltonian, and the scattering operator $S_{ij}$ is constrained by: $S_{ik}S_{jk}\dg= \delta_{ij} I$ and $S_{ki}\dg S_{kj} = \delta_{ij} I$ (see \cite[
 Appendix A]{GouGohYan08} and \cite[Sec. IV]{GouJam09} and the references therein for more details on multi-mode QSDEs). Note that the subscript $t$ on the quantum noises has been dropped for notational compactness in favor of the mode labels $\{i,j\} $. The multi-mode quantum noise increments are defined:
	\begin{align}\label{a2mode_flow}
		\int_t^{t + dt} ds \, b_i(s) \mapsto dB_i,\,\, {\rm and} \int_t^{t + d t} ds \, b_i^\dag(s) b_j(s) \mapsto d\Lambda_{ij}.
	\end{align}

%%% APPENDIX B:  WHITE NOISE FOR FOCK STATES %%%

\section{Quantum stochastic calculus for Fock states}

\subsection{Action of the quantum noise increments on Fock states}\label{qnoise_inc}
Recall the single photon state is defined by $\ket{1_{\xi}} = \int ds \ \xi(s) b\dg(s)\ket{0}  \equiv B\dg (\xi) \ket{0}.$ 
Acting the quantum noise increment $dB_t$ on this state gives
	\begin{align}
 		dB_t\ket{1_{\xi}} & =\int_t^{t + dt} dr \, b(r) \int ds  \,  \xi(s) b\dg(s)\ket{0} \nonumber \\
		&=\int_t^{t + dt}  \int dr ds  \,  \big( b\dg(s) b(r) +\delta(s-r) \big )\xi(s)   \ket{0} \nonumber \\
		&=\int_t^{t + dt}ds  \,  \xi(s)    \ket{0} \nn \\
		&= dt \,  \xi(t) \ket{0}. 
	\end{align}
Some of this algebraic manipulation can be simplified by using the Gardiner-Collett heuristic $dB_t\equiv dt\, b(t)$ \cite{GarZol00}.  Using this and the commutation relation $[b(t), B^\dagger(\xi)] = \xi(t)$ this procedure is extended incrementally to higher photon numbers.
%\begin{align}
%		 dB_t\ket{2_{\xi}} & = \frac{1}{\sqrt{2}} dt b(t) B^\dagger(\xi) B^\dagger(\xi) \ket{0} \nn \\
%		&= dt \sqrt{2} \xi(t)  \ket{1_\xi} .
%\end{align}
Through induction we obtain,
\begin{align}\label{appdB}
dB_{t}\ket{n_{\xi}}& =dt \sqrt{n} \xi(t) \ket{n-1_\xi}.
\end{align}
By the same procedure we find the action of $d\Lambda_t$,
\begin{align}
 d\Lambda_{t}\ket{n_{\xi}}& = dB\dg_t \sqrt{n} \xi(t) \ket{n-1_\xi}.
\end{align}

\subsection{Fock and $N$-photon It\={o} tables}\label{itotable}
The vacuum It\={o} table, \erf{Eq::ItoTableAppx}, can require modification for non-vacuum fields, such as thermal, coherent, and squeezed fields \cite{GarZol00,WisMilBook}.  Here we show, surprisingly, that the It\={o} tables for continuous-mode Fock states  and $N$-photon states are identical to the vacuum It\={o} table. This property was derived by the authors of Refs.~\cite{GJNphoton,GJNCgen} for a single photon, although never explicitly written down in those papers \cite{GJNpriv_comm}. 

Consider the expectation of $dB_t\, dB_t\dg$ for a single-photon Fock state. Normally ordering and simplifying gives
\begin{align}\label{Eq::itotab22}
 \expt{1_\xi|dB_tdB_t\dg|1_\xi} & =\expt{1_\xi| (dB_t\dg dB_t + dt) |1_\xi} = dt.
\end{align}	
%	Now we generalize table (\ref{Eq::FockStateItoTable}) to Fock states. First consider the single photon expectation of the product $dB_t\, dB_t\dg$
%\begin{align}\label{Eq::itotab1}
%		 \expt{1_\xi|dB_tdB_t\dg|1_\xi} & =\expt{1_\xi| \int_t^{t + dt} dr\,b(r) \int_t^{t + dt}ds\,b(s) |1_\xi}
%\end{align}
%Normally ordering the $dB_t\, dB_t\dg$ argument of \erf{Eq::itotab1}  and then simplifying gives
%\begin{align}\label{Eq::itotab2}
%		 \expt{1_\xi|dB_tdB_t\dg|1_\xi} & =\expt{1_\xi| (dB_t\dg dB_t + dt) |1_\xi} \nonumber \\
%		&=dt.
%\end{align}
Alone, \erf{Eq::itotab22} is not enough to specify the It\={o} rule rule for for $dB_t\, dB_t\dg$ because the action of the noise increments on Fock states couple different photon numbers, as in \erf{appdB}.  Consequently, we must consider cross expectations.  Only after showing that $ \expt{1_\xi|dB_tdB_t\dg|1_\xi}$, $ \expt{0|dB_tdB_t\dg|1_\xi}$, $ \expt{1_\xi |dB_tdB_t\dg|0}$, and $ \expt{0|dB_tdB_t\dg|0}$ are proportional to 0 or $dt$ can we say that $dB_t\, dB_t\dg = dt$ for the single photon It\={o} table.

	Now consider Fock states.  One must show that $ \expt{m_\xi|dB_tdB_t\dg|n_\xi}= \delta_{m,n}dt $ for all $m$ and $n$. Thankfully it is straightforward to show that after normally ordering the operators -- $dB_t, dB_t\dg, B(\xi)$, and $B\dg(\xi)$ -- %and repeatedly commuting the annihilation and creation operators (from the quantum noises) through those of the states
the only surviving term is proportional to $dt$ (terms proportional to $dt^2$ are set to zero).  Repeating this prescription for every product of the quantum noise increments in \erf{Eq::itotab22}, one can show the equivalence of the Fock and vacuum It\={o} tables.

%In general one has to show that all possible products of the quantum noises under  $ \expt{m_\xi|\,.\, |n_\xi}$   give rise to the corresponding It\={o} products in table (\ref{Eq::itotab22}). 

%	General Fock states require $ \expt{m_\xi|dB_tdB_t\dg|n_\xi}=dt$ for all $m$ and $n$. The first step to show this is to normally order the field operators and then take the expectation.  The only term that survives is proportional to $dt$ (terms proportional to $dt^2$ are set to zero).  Each of the It\={o} products follows similarly, and the resulting rules for Fock states are identical to  \erf{Eq::ItoTableAppx}.

	The It\={o} table for an $N$-photon state with an arbitrary spectral distribution function (within the quasi-monochromatic approximation) is also identical to the vacuum table. This follows from the occupation number representation, presented in Appendix \ref{Appendix::multiphotonStates}, which relies on a decomposition in a basis of orthogonal Fock states, each of which respects the its own Fock It\={o} table.

%%% APPENDIX:  General Multi-Photon States %%%
\section{Occupation number representation for general $N$-photon states} \label{Appendix::multiphotonStates}

	Here we review the occupation number representation of a general $N$-photon state presented in Ref. \cite{Roh07}.  In one dimension and in a single mode, a general quasi-monochromatic $N$-photon state can be written as 
	\begin{align}\label{eq:lalala}
	\begin{split}
		\ket{\psi_N} ={}\int d\omega_1 & \dots d\omega_N  \, \tilde\psi(\omega_1,\dots,\omega_N) \\
		& \times b\dg(\omega_1)\dots b\dg(\omega_N) \ket{0}\,.%\\
	\end{split}
	\end{align}
In the time domain, this becomes
	\begin{align} \label{Eq::multiphotonQuantumState}
		\ket{\psi_N} = \int dt_1\dots dt_N  \,  {\psi}(t_1,\dots,t_N) b\dg(t_1)\dots b\dg(t_N) \ket{0}\,,
	\end{align}
where the temporal envelope $\psi(t_1,\dots,t_N)$ is the Fourier transform of $\tilde\psi(\omega_1,\dots,\omega_N)$ \cite{BlowLouden90}  . The temporal envelope is in general neither factorable nor symmetric in $t_k$.  It can be expanded in a set of complex-valued, orthonormal basis functions that satisfy $\int dt \, \xi_i^*(t) \xi_j(t) = \delta_{i,j}$,
	\begin{align} \label{Eq::BasisWavepackets}
		\psi(t_1,\dots,t_N) = \sum_{i_1,\dots,i_N} \lambda'_{i_1,\dots,i_N} \xi_{i_1}(t_1)...\xi_{i_N}(t_N).
	\end{align}
Each subscript runs over the labels for the basis functions, i.e. $i_k \in \{1, 2,\dots \}$.  The expansion coefficients are given by the projection of the temporal envelope onto the basis functions,
	\begin{align} \label{Eq::MPExpansionC}
		\lambda'_{\alpha,\beta...,\zeta} = \int  dt_1\dots dt_N  \,  \xi_{\alpha}^*(t_1)\dots\xi_{\zeta}^*(t_N) {\psi}
			(t_1,\dots,t_N).
	\end{align}
	Defining a creation operator for a single photon in basis mode $\xi_{\alpha}(t)$ as $ B\dg (\xi_{\alpha}) =\int  dt \, \xi_{\alpha}(t) b\dg(t)$, and using Eq. (\ref{Eq::multiphotonQuantumState}-\ref{Eq::MPExpansionC}), we write the $N$-photon state as
	\begin{align}
		\ket{\psi_N} = \sum_{i_1,\dots,i_N} \lambda'_{i_1,\dots,i_N} B\dg(\xi_{i_1}) \dots B\dg(\xi_{i_N}) \ket{0}. 
	\end{align}
Acting these operators on vacuum yields an expression for the $N$-photon state in terms of basis Fock states, \erf{Eq::ContModeFockState}, in the basis functions,
	\begin{align} \label{Eq::FirstOcRep}
		\ket{\psi_N} = \sum_{i_1,\dots,i_N} \lambda'_{i_1,\dots,i_N} \sqrt{n_1! n_2! \dots} \ket{ {n_1}_{\xi_1}} \ket{{n_2}_		{\xi_2} } ...
	\end{align}
	Counting the number of subscripts of $\lambda'$ gives the total photon number $N$, which can be distributed among the basis Fock states in \erf{Eq::FirstOcRep}.  The number of photons $n_\alpha$ in a particular basis function $\xi_\alpha(t)$ is found by counting the number of indices of $\lambda'$ that are equal to $\alpha$.  For example, since they have 3 indices, the coefficients $\{ \lambda_{i_1,i_2,i_3}' \}$ all describe a 3-photon state.  The coefficient $\lambda_{1,1,4}'$ refers to the state $ \ket{ {2}_{\xi_1}} \ket{ {1}_{\xi_4}} $, in which the first and second photons are in $\xi_1(t)$ and the third in $\xi_4(t)$.  Due to the indistinguishability of photons, $\lambda_{1,4,1}'$ and $\lambda_{4,1,1}'$ are also coefficients for the state $ \ket{ {2}_{\xi_1}} \ket{ {1}_{\xi_4}} $, although they need not have the same value.  In general, $\lambda_{\alpha,\dots,\zeta}'$ is not invariant under permutation of its indices. The degree to which index-permutations are equal specifies the level of symmetry in the temporal envelope $\psi(t_1,...,t_N)$ \cite{Ou06, Ou08}.

Following \cite{Roh07}, we define a new set of coefficients
	\begin{align} \label{Eq::MPAmplitudes}
		\lambda_{i_1, \dots , i_N} = \sqrt{n_1!n_2!\dots} \sum_{\sigma \in \mathcal{S}_N} \lambda'_{\sigma( i_1,\dots,i_N ) }
	\end{align}
that sum over all permutations $\sigma$ (in the symmetric group $\mathcal{S}_N$) of the indices of coefficients of the type in \erf{Eq::MPExpansionC} so that no two coefficients in \erf{Eq::MPAmplitudes} refer to the same basis Fock state. The $N$-photon state of \erf{Eq::multiphotonQuantumState}, written in terms of these coefficients, is
	\begin{align} \label{AEq::NPhotonState}
		\ket{\psi_N} = \sum_{i_1 \leq \dots \leq i_N} \lambda_{i_1,  \dots,  i_N} \ket{ {n_1}_{\xi_1}} \ket{ {n_2}_{\xi_2}}	\dots
	\end{align}
Now it is clear that these algebraic acrobatics have culminated in a set of expansion coefficients that are precisely probability amplitudes,
	\begin{align}
		\sum_{i_1 \leq \dots \leq i_N} |\lambda_{i_1,\dots,i_N}|^2 = 1,
	\end{align}
and \erf{AEq::NPhotonState} is the \emph{occupation number representation} of the general $N$-photon state in \erf{Eq::multiphotonQuantumState}.

%\section{General N-photon states in two modes} \label{Appendix::TwoModemultiphotonStates}
\section{Multi-mode expectations} \label{Appendix::MultiMultimode}

In this section we extend our formalism to a countable number of modes. First we define a multi-mode Fock state in $T$ modes:
	\begin{align}\label{Eq:multimodefock}
		\ket{N_\alpha^1;\dots; N_\omega^T} =\frac{1}{\sqrt{N^1! \dots N^T!}} B\dg(\alpha)^{N^1} \dots B\dg(\omega)^{N^T}\ket{0},	
	\end{align}	
where there are $N^1$ photons in the first mode with the envelope $\alpha(t)$ and $\int_0^\infty ds \, |\alpha(s)|^2=1$.

To derive multi-mode mode master equations we must introduce notation, {\em different} from the main text, for representing asymmetric expectations. We define the multi-mode asymmetric expectation to be
	\begin{align} \label{Eq::MultiModeGeneralDensityOps}
		&\mathbbm{E}_{m^1;\dots ;\,m^T}^{n^1\,;\dots ;\,n^T}[ X(t) ] \\
		& =\! \Tr_{\rm sys + field} \!\left[ \!\left( \rho_{\rm sys}\! \otimes \!\op{m_\alpha^1}{n_\alpha^1} \!\otimes\! \dots  \otimes \op{ m_\omega^T}{ n_\omega^T }\right)\dg\!\! X(t)\! \right] \nonumber  \\
		& \equiv \!\Tr_{\rm sys} \left[ \big \{\varrho_{m^1;\dots; m^T}^{n^1\,;\dots;\,n^T}(t) \big\}\dg  X \right] ,\nn
	\end{align}
where the superscripts $n^1$ and $m^1$ on $\mathbb{E}[.]$ (and $\varrho$) refer to ``reference states'' in mode one. Note that \erf{Eq::MultiModeGeneralDensityOps} also defines the generalized multi-mode density operators $\varrho_{m^1;\dots ; m^T}^{n^1\,;\dots ;\,n^T}(t)$.

The final ingredient needed to derive the multi-mode mode master equation is the action of the quantum noise increments on Fock states:
	\begin{align}
		dB_j\ket{n_\alpha^1;\dots n_\omega^T} & = \,dt\sqrt{n^j} \theta(t) \ket{n_\alpha^1;\dots ;n-1^j_\theta;\dots n_\omega^T} ,\nn\\
		d\Lambda_{ij}\ket{n_\alpha^1;\dots n_\omega^T} & = dB_i\dg \sqrt{n^j} \theta(t) \ket{n_\alpha^1;\dots ;n-1^j_\theta;\dots n_\omega^T}.\nn
	\end{align}

%\section{Notational and Textual Conventions} 
%	
%	\subsection{Words in text}
%		Hyphenation:  $N$-photon continuous-mode wavepacket.  
%		
%		Capitalize Fock, Gaussian, Markovian.  
%		
%	\subsection{Equations}
%
%	Fock states have the wavepacket designation inside the ket.  Further, capital letters are reserved for the physical Fock state in question $\ket{N_\xi}$, while lowercase are for variables that refer to reference Fock states $\ket{n_\xi}$. 
%	
%	The wavepacket is denoted with the time indicator $\xi(t)$ unless its inside a ket $\ket{n_\xi}$ or part of a wavepacket creation operator $B\dg(\xi)$, which as you notice uses parentheses.
%
%	The quantum noise increments and stochastic processes on the field are denoted with a subscript $t$, $\Lambda_t$.  Stochastic processes on the system do not have this subscript and are denoted simply, $X$.  The general stochastic processes in appendices use the subscript $t$.
%	
%		Expectation values use brackets $\mathbbm{E}[O]$, partial expectations use parentheses, $\varpi(O)$.  
%		
%		In master equations (Heisenberg and Schrodinger), the term to the left of the equals sign has an explicit $t$-dependence, to the right of the equals all the terms do not.
%		\begin{align}
%			\frac{d}{dt} \varrho(t) = \varrho
%		\end{align}
%However, sometimes there is an overdot (as in the 2 photon example) rather than $d/dt$ in order to save space.
%		
%Integrals taken to a power use brackets,
%	\begin{align}
%		\left[ \int ds f(s)  \right]^3
%	\end{align}
%and the infinitessimals should come first.

%%% BIBLIOGRAPHY %%%	

\bibliographystyle{h-physrev3}

\begin{thebibliography}{99}

\bibitem{GioLor11}
V. Giovannetti, S. Lloyd, and L. Maccone, Nat. Photon. {\bf 4}, 222 (2011).
       %Journal = {Nat Photon},
       %Month = {04},
       %Number = {4},
       %Pages = {222--229},
       %Title = {Advances in quantum metrology},
       %Volume = {5}(2011)

\bibitem{Leroux11}
I. D. Leroux, M. H. Schleier-Smith, H. Zhang, and V. Vuleti\'{c}, Phys. Rev. A {\bf85}, 013803 (2012).

\bibitem{BevGrangier02}
A. Beveratos, R. Brouri, T. Gacoin, A. Villing, J-P. Poizat, and P. Grangier, Phys. Rev. Lett. {\bf 89}, 187901 (2002).
% title = {Single Photon Quantum Cryptography},

\bibitem{Kimble08}
H. J. Kimble,
% {The quantum internet},
Nature {\bf 453}, 1023 (2008).

\bibitem{MoeMauOlm07}
D. L. Moehring, P. Maunz, S. Olmschenk, K. C. Younge, D. N. Matsukevich, L.-M. Duan, and C. Monroe, 
%Entanglement of single-atom quantum bits at a distance
Nature {\bf 449}, 68 (2007). 

\bibitem{Agh11}
D. Aghamalyan and Y. Malakyan, Phys. Rev. A {\bf 84} 042305 (2011).

\bibitem{KLM01}
% �A scheme for efficient quantum computation with linear optics��, 
E. Knill, R. Laflamme, and G. J. Milburn, Nature {\bf 409}, 46 (2001).

\bibitem{Nielsen05}
P. P. Rohde, T. C. Ralph, and M. A. Nielsen, Phys. Rev. A {\bf 72}, 052332 (2005).

\bibitem{GarChi_book08}
J. C. Garrison, R. Y. Chiao, {\em Quantum Optics}, (Oxford University Press, Oxford, 2008).

\bibitem{Lou_book00}
R. Loudon, {\em The Quantum Theory of Light}, { third edition}, (Oxford University Press, Oxford, 2000).

\bibitem{Roh07}
%Spectral structure and decompositions of optical states, and their applications
P. Rohde, W. Mauerer, and C. Silberhorn, New J. Phys. {\bf 9}, 91 (2007).

\bibitem{BlowLouden90}
K. J. Blow, R. Loudon, and S. J. D. Phoenix, and T. J. Shepherd, Phys. Rev. A {\bf 42}, 4102 (1990).

\bibitem{BullColl10}
G. S. Buller and R. J. Collins, Meas. Sci. Technol. {\bf 21}, 012002 (2010).
% Single-photon generation and detection

\bibitem{Varcoe04}
B. Varcoe, S. Brattke, and H. Walther, New J. Phys. {\bf 6}, 97 (2004).

\bibitem{McKBocBoo04}
%  title={Deterministic generation of single photons from one atom trapped in a cavity},
 J. McKeever, A. Boca, A. D. Boozer, R. Miller, J. R. Buck, A. Kuzmich, and H. J. Kimble, {Science} {\bf 303}, 1992 (2004).
 
\bibitem{Yamamoto06}
E. Waks, E. Diamanti, and Y. Yamamoto, New J. Phys. {\bf 8}, 4 (2006).

\bibitem{Zeilinger06}
K. Sanaka, K. J. Resch, and A. Zeilinger, Phys. Rev. Lett. {\bf 96}, 083601 (2006).

\bibitem{KolBelDu08}
%Electro-Optic Modulation of Single Photons
P. Kolchin, C. Belthangady, S. Du, G. Y. Yin, and S. E. Harris, Phys. Rev. Lett. {\bf 101}, 103601 (2008).

\bibitem{Walmsley08}
P. J. Mosley, J. S. Lundeen, B. J. Smith, P. Wasylczyk, A. B. U'Ren, C. Silberhorn, and I. A. Walmsley, Phys. Rev. Lett. {\bf 100}, 133601 (2008).

\bibitem{Belthangady09}
S. Du, J. Wen, and C. Belthangady, Phys. Rev. A {\bf 79}, 043811 (2009). 

\bibitem{SpecBocMuc09}
%Phase shaping of single-photon wave packets
H. P. Specht, J. Bochmann, M. M\"ucke, B. Weber, E. Figueroa, D. L. Moehring, and G. Rempe, Nat. Photon. {\bf 3}, 469 (2009). 

\bibitem{Silberberg10}
I. Afek, O. Ambar, and Y. Silberberg, Science {\bf 328}, 5980 (2010).

\bibitem{Kuhn11}
P. B. R. Nisbet-Jones, J. Dilley, Daniel. Ljunggren, and A. Kuhn, New J. Phys. {\bf 13} 103036 (2011).

\bibitem{LeePatPar12}	
%Linear optical scheme for producing polarization-entangled NOON States
S. Lee, T. Paterek, H. S. Park, and H. Nha, Opt. Comm. {\bf 285}, 307 (2012).

\bibitem{GarPar94}
C. W. Gardiner and A. S. Parkins, Phys. Rev. A {\bf 50}, 1792 (1994).

\bibitem{GheEllPelZol99}
K. M. Gheri, K. Ellinger, T. Pellizzari, and P. Zoller, %{``Photon-Wavepackets as Flying Quantum Bits''}, in
%{\em Quantum computing: where do we want to go tomorrow?}, pg. {95}, (Vch Verlagsgesellschaft Mbh, 1999).
Fortschr. Phys. {\bf 46}, 401 (1998).

\bibitem{DomHorRit02}
P. Domokos, P. Horak, and H. Ritsch, Phys. Rev. A {\bf 65}, 033832 (2002).

\bibitem{DroHavBuz00}
  %{Stimulated emission via quantum interference: scattering of one-photon packets on an atom in a ground state}
  G. Drobn\`y, and M. Havukainen, and V. Bu\v{z}ek, J. Mod. Opt. {\bf 47}, 851 (2000).



\bibitem{SheFan05}
%Coherent photon transport from spontaneous emission in one-dimensional waveguides
J. T. Shen and S. Fan, Opt. Lett. {\bf 30}, 2001 (2005).

\bibitem{ZhoGonSun08}
%Controllable Scattering of a Single Photon inside a One-Dimensional Resonator Waveguide
L. Zhou, Z. R. Gong, Y. X. Liu, C. P. Sun, and F. Nori, Phys. Rev. Lett. {\bf 101}, 100501 (2008).

\bibitem{SheFan07b}
%Strongly Correlated Two-Photon Transport in a One-Dimensional Waveguide Coupled to a Two-Level System
J. T. Shen and S. Fan, Phys. Rev. Lett. {\bf 98}, 153003 (2007).  

\bibitem{Kos08}
%Single-photon filtering by a cavity quantum electrodynamics system
K. Koshino, Phys. Rev. A {\bf 77}, 023805 (2008).

\bibitem{LonSchBus09}
%Dynamics of photon transport through quantum impurities in dispersion-engineered one-dimensional systems
P. Longo, P. Schmitteckert, and K. Busch, J. Opt. A: Pure Appl. Opt. {\bf 11}, 114009 (2009). 

\bibitem{ZheGauBar10}
%Waveguide QED: Many-body bound-state effects in coherent and Fock-state scattering from a two-level system
H. Zheng, D. J. Gauthier, and H. U. Baranger, Phys. Rev. A {\bf 82}, 063816 (2010).
  
\bibitem{CheWubMor11} 
%Coherent single-photon absorption by single emitters coupled to one-dimensional nanophotonic waveguides
Y. Chen, M. Wubs, J. M\o rk, and A. F. Koenderink, New J. Phys. {\bf 13}, 103010 (2011).

\bibitem{Roy11}
% Correlated few-photon transport in one-dimensional waveguides: Linear and nonlinear dispersions
D. Roy, Phys. Rev. A, {\bf 83}, 043823 (2011).

\bibitem{Roy10}
%Few-photon optical diode
D. Roy, Phys. Rev. B {\bf 81}, 155117 (2010).

\bibitem{Ely12}
%Interaction of a single-photon wave packet with an excited atom
P. V. Elyutin, Phys. Rev. A {\bf 85}, 033816 (2012).

\bibitem{StoAlbLeu07}
M. Stobi\'nska, G. Alber, and G. Leuchs, 
%Perfect excitation of a matter qubit by a single photon in free space
EPL {\bf 86}, 14007 (2009).

\bibitem{WanSheSca10}
%Efficient excitation of a two-level atom by a single photon in a propagating mode
Y. Wang, J. Min\'ar, L. Sheridan, and V. Scarani, Phys. Rev. A {\bf 83}, 063842 (2011).

\bibitem{StoAlbLeu10}
M. Stobi\'nska, G. Alber, and G. Leuchs, {\em Chapter 8 - Quantum Electrodynamics of One-Photon Wave Packets} in {\em Unstable States in the Continuous Spectra, Part I: Analysis, Concepts, Methods, and Results}, Edited by Cleanthes A. Nicolaides and Erkki Br\"{a}ndas, Adv. Quant. Chem. {\bf 60}, pp. 457-483 (2010). Also available as arXiv:1002.3059.
%
%@incollection{Stobiãska2010457,
%title = "Chapter 8 - Quantum Electrodynamics of One-Photon Wave Packets",
%editor = "Cleanthes A. Nicolaides and Erkki BrÌ?ndas",
%booktitle = "Unstable States in the Continuous Spectra, Part I: Analysis, Concepts, Methods, and Results",
%publisher = "Academic Press",
%year = "2010",
%volume = "60",
%pages = "457 - 483",
%series = "Advances in Quantum Chemistry",
%issn = "0065-3276",
%doi = "10.1016/S0065-3276(10)60008-1",
%url = "http://www.sciencedirect.com/science/article/pii/S0065327610600081",
%author = "M. Stobiãska and G. Alber and G. Leuchs"
%}


\bibitem{RepSheFan10}
%Full inversion of a two-level atom with a single-photon pulse in one-dimensional geometries
E. Rephaeli, Jung-Tsung Shen, and S. Fan, Phys. Rev. A {\bf 82}, 033804 (2010).

\bibitem{YudRei08}
%Multiphoton scattering in a one-dimensional waveguide with resonant atoms
V. I. Yudson and P. Reineker, Phys. Rev. A {\bf 78}, 052713 (2008).

\bibitem{ColGar84}
M. J. Collett and C. W. Gardiner, Phys. Rev. A {\bf 30}, 1386 (1984).

\bibitem{GarCol85}
C. W. Gardiner and M. J. Collett, Phys. Rev. A {\bf 31}, 3761 (1985).

\bibitem{GarZol00}
C. W. Gardiner and P. Zoller, {\em Quantum noise} (Springer Berlin, 2000).

\bibitem{YurDen84}
%Quantum network theory
B. Yurke and J. S. Denker, Phys. Rev. A, {\bf 29} 1419 (1984).

\bibitem{Caves82}
C. M. Caves, Phys. Rev. D {\bf 26}, 1817 (1982).

\bibitem{qnoise_rmp10}
A. A. Clerk, M. H. Devoret, S. M. Girvin, F. Marquardt, and R. J. Schoelkopf, Rev. Mod. Phys. {\bf 82}, 1155 (2010).

\bibitem{DumParZol92}
R. Dum, A. S. Parkins, P. Zoller, and C. W. Gardiner, Phys. Rev. A {\bf 46}, 4382 (1992).

\bibitem{Gar93}
C. W. Gardiner, Phys. Rev. Lett. {\bf 70}, 2269 (1993).

\bibitem{Car93}
H. J. Carmichael, Phys. Rev. Lett. {\bf 70}, 2273 (1993).

\bibitem{GJNphoton}
% Filtering equations for fields in a single photon state
J. E. Gough, M. R. James, and H. I. Nurdin, in \emph{Proceedings of the joint 50th IEEE Conference on Decision and Control and European Control Conference (CDC-ECC), Orlando, 2011}, pp. 5570-5576. Also available as arXiv:1107.2973.

%author={Gough, John E. and James, Matthew R. and Nurdin, Hendra I.}, 
%booktitle={Decision and Control and European Control Conference (CDC-ECC), 2011 50th IEEE Conference on}, title={Quantum master equation and filter for systems driven by fields in a single photon state}, 
%year={2011}, 
%month={dec.}, 
%volume={}, 
%number={}, 
%pages={5570 -5576}, 
%keywords={}, 
%doi={10.1109/CDC.2011.6160675}, 
%ISSN={0743-1546},}



\bibitem{GJNCgen}
%Quantum Filtering (Quantum Trajectories) for Systems Driven by Fields in Single Photon States and Superposition of Coherent States
J. E. Gough, M. R. James, H. I. Nurdin, and J. Combes, arXiv:1107.2976.



\bibitem{Aoki09}
T. Aoki, A. S. Parkins, D. J. Alton, C. A. Regal, B. Dayan, E. Ostby, K. J. Vahala, and H. J. Kimble, Phys. Rev. Lett. {\bf 102} 083601 (2009).

\bibitem{Chang07}
D. Chang, A. S\o rensen, E. Demler, and M. Lukin, Nat. Phys. {\bf 3}, 807 (2007).

\bibitem{Spillane08}
S. M. Spillane, G. S. Pati, K. Salit, M. Hall, P. Kumar, R. G. Beausoleil, and M. S. Shahriar, Phys. Rev. Lett. {\bf 100}, 233602 (2008).

\bibitem{Vetsch10}
E. Vetsch, D. Reitz, G. Sagu\'{e}, R. Schmidt, S. T. Dawkins, and A. Rauschenbeutel, Phys. Rev. Lett. {\bf 104}, 203603 (2010).


\bibitem{vanHStoMab05a}
R. van Handel, J. K. Stockton, and H. Mabuchi, J. Opt. B: Quantum Semiclass. Opt. {\bf 7}, 179 (2005).

\bibitem{VanHove55}
%  title={Energy corrections and persistent perturbation effects in continuous spectra},
  L. Van Hove, Physica {\bf 21}, 901 (1955).
  
\bibitem{Accardi}
L. Accardi, Y. G. Lu, and I. Volovich, {\em Quantum Theory and its Stochastic Limit}, (Springer, 2002).

\bibitem{YanKim03a}
  %title={Transfer function approach to quantum control-part I: Dynamics of quantum feedback systems},
M. Yanagisawa and H. Kimura, IEEE Transactions on Automatic Control {\bf 48}, 2107 (2003).

\bibitem{Louisell1973book}
W.H. Louisell, {\em Quantum Statistical Properties of Radiation}, (Wiley, 1973).

\bibitem{Van11}
M. R. Vanner, Phys. Rev. X {\bf 1}, 021011 (2011).
%Selective Linear or Quadratic Optomechanical Coupling via Measurement

\bibitem{DeuJes09}
I. H. Deutsch and P. S. Jessen, Opt. Comm. {\bf283}, 681 (2009).

\bibitem{Mil07}
G. J. Milburn, in Springer Handbook of Lasers and Optics, edited by F. Tr\"{a}ger (Springer, 2007) Chap. 14, pp. 1053.

\bibitem{Mil08}
G. J. Milburn, Eur. Phys. J. Special Topics {\bf 159}, 113 (2008).

\bibitem{WisMilBook}
H. M. Wiseman and G. J. Milburn, \textsl{Quantum Measurement and Control}, (Cambridge Univ. Press, Cambridge, 2010).

\bibitem{RohdeSep06}
%Quantum state tomography of single photons in the spectral degree of freedom
P. Rohde, arXiv:quant-ph/0609005v1.

\bibitem{WasKolRob07}
%Spectral Density Matrix of a Single Photon Measured
W. Wasilewski, P. Kolenderski, and R. Frankowski, Phys. Rev. Lett. {\bf 99}, 123601 (2007).

\bibitem{Ou06}
%Temporal distinguishability of an N-photon state and its characterization by quantum interference
Z. Y. Ou, Phys. Rev. A {\bf 74}, 063808 (2006).

\bibitem{SilDeu03}
A. Silberfarb and I. H. Deutsch, Phys. Rev. A {\bf 68}, 013817 (2003).

\bibitem{GouGohYan08}
% Linear quantum feedback networks
J. E. Gough, R. Gohm, and M. Yanagisawa, Phys. Rev. A {\bf 78}, 062104 (2008).

\bibitem{GouJam09}
  %The Series Product and Its Application to Quantum Feedforward and Feedback Networks
J. Gough and M. R. James, IEEE Transactions on Automatic Control {\bf 54}, 2530 (2009).

\bibitem{GouJam09b}
%Quantum Feedback Networks: Hamiltonian Formulation
J. Gough and M. R. James, Commun. Math. Phys. {\bf 287}, 1109 (2009).

\bibitem{DeuChi92}
% Biphoton bound states in an optical quantum wire
I. H. Deutsch, R. Y. Chiao, and J.C. Garrison, Phys. Rev. Lett. {\bf 69}, 3627 (1992).

\bibitem{CarBook93}
H. J. Carmichael, {\em An Open Systems Approach to Quantum Optics} (Springer: lecture notes in physics vol. 18, 1993).

\bibitem{BouvanHJam07}
L. Bouten, R. van Handel and M. R. James, SIAM Journal on Control and Optimization {\bf 46}, 2199 (2007).

\bibitem{BasAkrMil12}
%Phonon number measurements using single photon opto-mechanics
S. Basiri Esfahani, U. Akram, G. J. Milburn, arXiv:1205.3240.

\bibitem{YanKim03b}
  %title={Transfer function approach to quantum control-part Part II: Control Concepts and Applications
M. Yanagisawa and H. Kimura, IEEE Transactions on Automatic Control, {\bf 48} 2121 (2003).

\bibitem{GouJam10}
% Quantum Dissipative Systems and Feedback
J. Gough and M. R. James, IEEE Transactions on Automatic Control {\bf 55}, 1806 (2010).

\bibitem{Gou06}
%Quantum Stratonovich calculus and the quantum Wong-Zakai theorem
J. Gough, J. Math. Phys. {\bf 47}, 113509 (2006).

%\bibitem{BrePetBook02}
%H. P. Breuer and F. Petruccione, {\em The theory of open quantum systems}, (Oxford University Press, USA 2002).




\bibitem{HudPar84}
R. L. Hudson and K. R. Parthasarathy, Commun. Math. Phys. {\bf 93}, 301 (1984).

\bibitem{Par92}
K. R. Parthasarathy. {\em An Introduction to Quantum Stochastic Calculus} (Birkhauser, 1992).

\bibitem{Bar06}
 %A. Barchielli, %Continual Measurements in Quantum Mechanics and Quantum Stochastic Calculus.
 %I. S. Attal, A. Joye, C.-A. Pillet (eds.), Open Quantum Systems III, Lect.
%Notes Math. 1882 (Springer, Berlin, 2006), pp. 207.%{291; ISBN: 978-3-540-30993-2; DOI:10.1007/b128453.

A. Barchielli in \emph{Open Quantum Systems III: Recent Developments (Lecture Notes in Mathematics)}, edited by I. S. Attal, A. Joye, and C.-A. Pillet (Springer, Berlin, 2006), p. 207.

\bibitem{GoughvHan07}
J. Gough and R. van Handel, J. Stat. Phys. {\bf127}, 3 (2007).

\bibitem{GJNpriv_comm}
J.E. Gough, M.R. James, and H.I. Nurdin, private communication (2011).

\bibitem{Ou08}
%Characterizing temporal distinguishability of an N-photon state by a generalized photon bunching effect with multiphoton interference
Z. Y. Ou, Phys. Rev. A {\bf 77}, 043829 (2008).

%\bibitem{mikeike} M. A. Nielsen and I. L. Chuang, \textit{Quantum Computation and Quantum Information}, Cambridge University Press, Cambridge, England, 2000.



\end{thebibliography}

\end{document}